\openin 1 lanlmac
\ifeof 1 
  \message{[Load harvmac]}
  \input harvmac 
\else 
  \message{[Load lanlmac]}
  \input lanlmac
\fi
\closein 1
\input amssym
\input epsf
\input tables
\noblackbox

\newif\ifhypertex
\ifx\hyperdef\UnDeFiNeD
    \hypertexfalse
    \message{[HYPERTEX MODE OFF]}
    \def\hyperref#1#2#3#4{#4}
    \def\hyperdef#1#2#3#4{#4}
    \def\hypernoname{}
    \def\e@tf@ur#1{}
    
\else
    \hypertextrue
    \message{[HYPERTEX MODE ON]}
    
\fi

\ifx\answ\bigans

\magnification=1200\baselineskip=15pt plus 2pt minus 1pt
%
\advance\voffset by.6truecm
\hsize=6.15truein\vsize=600.truept\hsbody=\hsize\hstitle=\hsize
\else\let\lr=L

\magnification=1000\baselineskip=15pt plus 2pt minus 1pt
%
\hoffset=-0.75truein\voffset=-.0truein
\vsize=6.5truein
\hstitle=8.truein\hsbody=4.75truein
\fullhsize=10truein\hsize=\hsbody
\fi
\parskip=4pt plus 15pt minus 1pt

\newif\iffigureexists
\def\checkex#1{
\relax
\openin 1 #1
   \ifeof 1
      \figureexistsfalse
      \immediate\write20{FIGURE FILE #1 NOT FOUND}
   \else 
      \figureexiststrue
   \fi 
\closein 1}
\def\missbox#1#2{$\vcenter{\hrule
\hbox{\vrule height#1\kern1.truein
\raise.5truein\hbox{#2} \kern1.truein \vrule} \hrule}$}
\def\lfig#1{
\let\labelflag=#1%
\def\numb@rone{#1}%
\ifx\labelflag\UnDeFiNeD%
{\xdef#1{\the\figno}%
\writedef{#1\leftbracket{\the\figno}}%
\global\advance\figno by1%
}\fi{\hyperref{}{figure}{{\numb@rone}}{Fig.~{\numb@rone}}}}
\def\figinsert#1#2#3#4{
\checkex{#4}
\def\figsize{#3}%
\let\flag=#1\ifx\flag\UnDeFiNeD
{\xdef#1{\the\figno}%
\writedef{#1\leftbracket{\the\figno}}%
\global\advance\figno by1%
}\fi
\goodbreak
\midinsert
  \iffigureexists
     \centerline{\epsfysize\figsize\epsfbox{#4}}%
  \else
     \vskip.05truein
     \centerline{\missbox\figsize{#4 not found!}}%
     \vskip.05truein
  \fi
{\smallskip%
\leftskip 4pc \rightskip 4pc%
\noindent\ninepoint\sl \baselineskip=11pt%
{\bf{\hyperdef\hypernoname{figure}{{#1}}{Fig.~{#1}}}:~}#2%
\smallskip}\bigskip\endinsert%
}

\newcount\tabno
\tabno=1
\def\ltab#1{
\let\labelflag=#1%
\def\numb@rone{#1}%
\ifx\labelflag\UnDeFiNeD{%
  \xdef#1{\the\tabno}%
  \writedef{#1\leftbracket{\the\tabno}}%
  \global\advance\tabno by1%
}%
\fi%
{\hyperref{}{table}{{\numb@rone}}{Table~{\numb@rone}}}}
\def\tabinsert#1#2#3{
\let\flag=#1
\ifx\flag\UnDeFiNeD
  {\xdef#1{\the\tabno}
   \writedef{#1\leftbracket{\the\tabno}}
   \global\advance\tabno by1 }
\fi
\vbox{\bigskip #3 \smallskip
\leftskip 4pc \rightskip 4pc
\noindent\ninepoint\sl \baselineskip=11pt
{\bf{\hyperdef\hypernoname{table}{{#1}}{Table~{#1}}}.~}#2
\smallskip}
\bigskip}

\def\ie{{\it i.e.\ }}
\def\eg{{\it e.g.\ }}
\def\cf{{\it cf.\ }}
\def\frac#1#2{{#1 \over #2}}
\def\llangle{\langle\!\!\left\langle}
\def\rrangle{\rangle\!\!\right\rangle}
\def\corr#1{\left\langle {#1} \right\rangle}
\def\norm#1{\left| {#1} \right|}
\def\Im#1{{{\rm Im}\,#1}}
\def\sbox{{\vrule height 6.5pt width 0.5pt depth 0.5pt%
           \hskip -0.5pt%
           \vrule height 0pt width 7pt depth 0.5pt%
           \hskip -7pt%
           \vrule height 6.5pt width 7pt depth -6pt%
           \hskip -0.5pt%
           \vrule height 6.5pt width 0.5pt depth 0.5pt}}
\def\tbox{{\sbox\hskip-0.5pt\sbox\hskip-0.5pt\sbox}}
\def\IC{{\Bbb C}}
\def\IP{{\Bbb P}}
\def\IR{{\Bbb R}}
\def\IZ{{\Bbb Z}}

\def\cD{{\cal D}}
\def\cI{{\cal I}}
\def\cL{{\cal L}}
\def\cN{{\cal N}}
\def\cQ{{\cal Q}}
\def\cT{{\cal T}}
\def\cW{{\cal W}}
\def\bxi{{\xi\hskip -1.03ex \xi\hskip -1.03ex\xi}}
\def\qf{q_{\rm f}}            
\def\pf{\wp}                  
\def\cmat#1{{\underline{#1}}} 
\def\cfa{{\frak c}}           
\def\gf{{\cal C}}             




\lref\AntoniadisPM{
I.~Antoniadis, E.~Gava, K.~S.~Narain and T.~R.~Taylor,
``Superstring threshold corrections to Yukawa couplings,''
Nucl.\ Phys.\ B {\bf 407}, 706 (1993)
[arXiv:hep-th/9212045].
}

\lref\AshokZB{
S.~K.~Ashok, E.~Dell'Aquila and D.~E.~Diaconescu,
``Fractional branes in Landau-Ginzburg orbifolds,''
Adv.\ Theor.\ Math.\ Phys.\  {\bf 8}, 461 (2004)
[arXiv:hep-th/0401135].
}

\lref\AshokXQ{
S.~K.~Ashok, E.~Dell'Aquila, D.~E.~Diaconescu and B.~Florea,
``Obstructed D-branes in Landau-Ginzburg orbifolds,''
Adv.\ Theor.\ Math.\ Phys.\  {\bf 8}, 427 (2004)
[arXiv:hep-th/0404167].
}

\lref\AspinwallBS{
P.~S.~Aspinwall and S.~Katz,
``Computation Of Superpotentials For D-Branes,''
Commun.\ Math.\ Phys.\  {\bf 264}, 227 (2006)
[arXiv:hep-th/0412209].
}

\lref\BershadskyCX{
M.~Bershadsky, S.~Cecotti, H.~Ooguri and C.~Vafa,
``Kodaira-Spencer theory of gravity and exact results for quantum string
amplitudes,''
Commun.\ Math.\ Phys.\  {\bf 165}, 311 (1994)
[arXiv:hep-th/9309140].
}

\lref\BrunnerJQ{
I.~Brunner, M.~R.~Douglas, A.~E.~Lawrence and C.~R\"omelsberger,
``D-branes on the quintic,''
JHEP {\bf 0008}, 015 (2000)
[arXiv:hep-th/9906200].
}

\lref\BrunnerDC{
I.~Brunner, M.~Herbst, W.~Lerche and B.~Scheuner,
``Landau-Ginzburg realization of open string TFT,''
arXiv:hep-th/0305133.
}

\lref\BrunnerMT{
I.~Brunner, M.~Herbst, W.~Lerche and J.~Walcher,
``Matrix factorizations and mirror symmetry: The cubic curve,''
arXiv:hep-th/0408243.
}

\lref\CremadesQJ{
D.~Cremades, L.~E.~Ib\'a\~nez and F.~Marchesano,
``Yukawa couplings in intersecting D-brane models,''
JHEP {\bf 0307}, 038 (2003)
[arXiv:hep-th/0302105].
}

\lref\DijkgraafDJ{
R.~Dijkgraaf, H.~L.~Verlinde and E.~P.~Verlinde,
``Topological Strings in $d<1$,''
Nucl.\ Phys.\ B {\bf 352}, 59 (1991).
}

\lref\DijkgraafFC{
R.~Dijkgraaf and C.~Vafa,
``Matrix models, topological strings, and supersymmetric gauge theories,''
Nucl.\ Phys.\ B {\bf 644}, 3 (2002)
[arXiv:hep-th/0206255].
}

\lref\DijkgraafVW{
R.~Dijkgraaf and C.~Vafa,
``On geometry and matrix models,''
Nucl.\ Phys.\ B {\bf 644}, 21 (2002)
[arXiv:hep-th/0207106].
}

\lref\DouglasQW{
M.~R.~Douglas, B.~Fiol and C.~R\"omelsberger,
``The spectrum of BPS branes on a noncompact Calabi-Yau,''
JHEP {\bf 0509}, 057 (2005)
[arXiv:hep-th/0003263].
}

\lref\DouglasFR{
M.~R.~Douglas, S.~Govindarajan, T.~Jayaraman and A.~Tomasiello,
``D-branes on Calabi-Yau manifolds and superpotentials,''
Commun.\ Math.\ Phys.\  {\bf 248}, 85 (2004)
[arXiv:hep-th/0203173].
}

\lref\EzhuthachanJR{
B.~Ezhuthachan, S.~Govindarajan and T.~Jayaraman,
``A quantum McKay correspondence for fractional 2p-branes on LG  orbifolds,''
JHEP {\bf 0508}, 050 (2005)
[arXiv:hep-th/0504164].
}

\lref\EzhuthachanGU{
B.~Ezhuthachan, S.~Govindarajan and T.~Jayaraman,
``Fractional two-branes, toric orbifolds and the quantum McKay
correspondence,''
arXiv:hep-th/0606154.
}

\lref\GepnerVZ{
D.~Gepner,
``Exactly Solvable String Compactifications On Manifolds Of SU(N) Holonomy,''
Phys.\ Lett.\ B {\bf 199}, 380 (1987).
}

\lref\GepnerQI{
D.~Gepner,
``Space-Time Supersymmetry In Compactified String Theory And Superconformal
Models,''
Nucl.\ Phys.\ B {\bf 296}, 757 (1988).
}

\lref\GovindarajanJS{
S.~Govindarajan, T.~Jayaraman and T.~Sarkar,
``Worldsheet approaches to D-branes on supersymmetric cycles,''
Nucl.\ Phys.\ B {\bf 580}, 519 (2000)
[arXiv:hep-th/9907131].
}
	     
\lref\GovindarajanKR{
S.~Govindarajan and T.~Jayaraman,
``Boundary fermions, coherent sheaves and D-branes on Calabi-Yau
manifolds,''
Nucl.\ Phys.\ B {\bf 618}, 50 (2001)
[arXiv:hep-th/0104126].
}
	     
\lref\GovindarajanIM{
S.~Govindarajan, H.~Jockers, W.~Lerche and N.~P.~Warner,
``Tachyon condensation on the elliptic curve,''
arXiv:hep-th/0512208.
}

\lref\GreeneUT{
B.~R.~Greene, C.~Vafa and N.~P.~Warner,
``Calabi-Yau manifolds and renormalization group flows,''
Nucl.\ Phys.\ B {\bf 324}, 371 (1989).
}

\lref\HellermanBU{
S.~Hellerman, S.~Kachru, A.~E.~Lawrence and J.~McGreevy,
``Linear sigma models for open strings,''
JHEP {\bf 0207}, 002 (2002)
[arXiv:hep-th/0109069].
}

\lref\HerbstAX{
M.~Herbst and C.~I.~Lazaroiu,
``Localization and traces in open-closed topological Landau-Ginzburg
models,''
JHEP {\bf 0505}, 044 (2005)
[arXiv:hep-th/0404184].
}

\lref\HerbstJP{
M.~Herbst, C.~I.~Lazaroiu and W.~Lerche,
``Superpotentials, A(infinity) relations and WDVV equations for open
topological strings,''
JHEP {\bf 0502}, 071 (2005)
[arXiv:hep-th/0402110].
}

\lref\HerbstKT{
  M.~Herbst,
  ``Quantum A-infinity structures for open-closed topological strings,''
  arXiv:hep-th/0602018.
}

\lref\HerbstNN{
M.~Herbst, W.~Lerche and D.~Nemeschansky,
``Instanton geometry and quantum A(infinity) structure on the elliptic
curve,''
arXiv:hep-th/0603085.
}

\lref\HerbstZM{
M.~Herbst, C.~I.~Lazaroiu and W.~Lerche,
``D-brane effective action and tachyon condensation in topological minimal
models,''
JHEP {\bf 0503}, 078 (2005)
[arXiv:hep-th/0405138].
}

\lref\HofmanCE{
C.~Hofman and W.~K.~Ma,
``Deformations of topological open strings,''
JHEP {\bf 0101}, 035 (2001)
[arXiv:hep-th/0006120].
}

\lref\HoriCK{
K.~Hori, A.~Iqbal and C.~Vafa,
``D-branes and mirror symmetry,''
arXiv:hep-th/0005247.
}

\lref\HoriJA{
K.~Hori and J.~Walcher,
``F-term equations near Gepner points,''
JHEP {\bf 0501}, 008 (2005)
[arXiv:hep-th/0404196].
}

\lref\KachruPG{
S.~Kachru and E.~Witten,
``Computing the complete massless spectrum of a Landau-Ginzburg orbifold,''
Nucl.\ Phys.\ B {\bf 407}, 637 (1993)
[arXiv:hep-th/9307038].
}

\lref\KachruAN{
S.~Kachru, S.~Katz, A.~E.~Lawrence and J.~McGreevy,
``Mirror symmetry for open strings,''
Phys.\ Rev.\ D {\bf 62}, 126005 (2000)
[arXiv:hep-th/0006047].
}

\lref\KachruIH{
S.~Kachru, S.~Katz, A.~E.~Lawrence and J.~McGreevy,
``Open string instantons and superpotentials,''
Phys.\ Rev.\ D {\bf 62}, 026001 (2000)
[arXiv:hep-th/9912151].
}

\lref\KapustinBI{
A.~Kapustin and Y.~Li,
``D-branes in Landau-Ginzburg models and algebraic geometry,''
JHEP {\bf 0312}, 005 (2003)
[arXiv:hep-th/0210296].
}

\lref\KapustinRC{
A.~Kapustin and Y.~Li,
``D-branes in topological minimal models: The Landau-Ginzburg approach,''
JHEP {\bf 0407}, 045 (2004)
[arXiv:hep-th/0306001].
}

\lref\KnappRD{
J.~Knapp and H.~Omer,
``Matrix factorizations, minimal models and Massey products,''
JHEP {\bf 0605}, 064 (2006)
[arXiv:hep-th/0604189].
}

\lref\KutasovAQ{
D.~Kutasov, M.~Mari\~no and G.~W.~Moore,
``Remarks on tachyon condensation in superstring field theory,''
arXiv:hep-th/0010108.
}

\lref\LazaroiuZI{
C.~I.~Lazaroiu,
``On the boundary coupling of topological Landau-Ginzburg models,''
JHEP {\bf 0505}, 037 (2005)
[arXiv:hep-th/0312286].
}

\lref\LercheCS{
W.~Lerche, D.~L\"ust and N.~P.~Warner,
``Duality Symmetries In N=2 Landau-Ginzburg Models,''
Phys.\ Lett.\ B {\bf 231}, 417 (1989).
}

\lref\LercheWM{
W.~Lerche, D.~J.~Smit and N.~P.~Warner,
``Differential equations for periods and flat coordinates in two-dimensional
topological matter theories,''
Nucl.\ Phys.\ B {\bf 372}, 87 (1992)
[arXiv:hep-th/9108013].
}

\lref\MartinecZU{
E.~J.~Martinec,
``Algebraic geometry and effective Lagrangians,''
Phys.\ Lett.\ B {\bf 217}, 431 (1989).
}

\lref\NiarchosSI{
V.~Niarchos and N.~Prezas,
``Boundary superstring field theory,''
Nucl.\ Phys.\ B {\bf 619}, 51 (2001)
[arXiv:hep-th/0103102].
}

\lref\OoguriBV{
H.~Ooguri and C.~Vafa,
``Knot invariants and topological strings,''
Nucl.\ Phys.\ B {\bf 577}, 419 (2000)
[arXiv:hep-th/9912123].
}

\lref\OoguriTT{
H.~Ooguri and C.~Vafa,
``Gravity induced C-deformation,''
Adv.\ Theor.\ Math.\ Phys.\  {\bf 7}, 405 (2004)
[arXiv:hep-th/0303063].
}

\lref\PolishchukDB{
A.~Polishchuk and E.~Zaslow,
``Categorical mirror symmetry: The Elliptic curve,''
Adv.\ Theor.\ Math.\ Phys.\  {\bf 2}, 443 (1998)
[arXiv:math.ag/9801119].
}

\lref\SchellekensXH{
A.~N.~Schellekens and N.~P.~Warner,
``Anomalies, Characters And Strings,''
Nucl.\ Phys.\ B {\bf 287}, 317 (1987).
}

\lref\ShatashviliKK{
S.~L.~Shatashvili,
``Comment on the background independent open string theory,''
Phys.\ Lett.\ B {\bf 311}, 83 (1993)
[arXiv:hep-th/9303143].
}

\lref\SiqvelandAA{
A.~Siqveland, 
``The Method of Computing Formal Moduli,'' 
J.\ Alg.\ {\bf 241}, 292--327 (2001).
}

\lref\SiqvelandBB{
A.~Siqveland, 
``Global Matric Massey Products and the Compactified Jacobian of the E6--Singularity,'' 
J.\ Alg.\ {\bf 241} 259--291 (2001).
}

\lref\SiqvelandCC{
A.~Siqveland, 
``Generalized Matric Massey Products for Graded Modules,'' 
arXiv:math.AG/0603425.
}

\lref\VafaMU{
C.~Vafa,
``Topological Landau-Ginzburg models,''
Mod.\ Phys.\ Lett.\ A {\bf 6}, 337 (1991).
}

\lref\VafaWI{
C.~Vafa,
``Superstrings and topological strings at large N,''
J.\ Math.\ Phys.\  {\bf 42}, 2798 (2001)
[arXiv:hep-th/0008142].
}

\lref\WalcherTX{
J.~Walcher,
``Stability of Landau-Ginzburg branes,''
J.\ Math.\ Phys.\  {\bf 46}, 082305 (2005)
[arXiv:hep-th/0412274].
}

\lref\WarnerAY{
N.~P.~Warner,
``Supersymmetry in boundary integrable models,''
Nucl.\ Phys.\ B {\bf 450}, 663 (1995)
[arXiv:hep-th/9506064].
}

\lref\WittenED{
E.~Witten,
``Quantum background independence in string theory,''
arXiv:hep-th/9306122.
}

\lref\WittenJG{
E.~Witten,
``On the Landau-Ginzburg description of N=2 minimal models,''
Int.\ J.\ Mod.\ Phys.\ A {\bf 9}, 4783 (1994)
[arXiv:hep-th/9304026].
}

\lref\WittenQY{
E.~Witten,
``On background independent open string field theory,''
Phys.\ Rev.\ D {\bf 46}, 5467 (1992)
[arXiv:hep-th/9208027].
}

\lref\WittenYC{
E.~Witten,
``Phases of N = 2 theories in two dimensions,''
Nucl.\ Phys.\ B {\bf 403}, 159 (1993)
[arXiv:hep-th/9301042].
}

\lref\WittenZZ{
E.~Witten,
``Mirror manifolds and topological field theory,''
arXiv:hep-th/9112056.
}


\lref\LiE{
M.~A.~A.~van~Leeuwen, A.~M.~Cohen and B.~Lisser,
``LiE~2.2.2: A computer algebra package,''
{\tt http://www-math.univ-poitiers.fr/\~{}maavl/LiE/}.
}


\Title{
\vbox{
\hbox{\tt hep-th/0608027}\vskip -.15cm
\hbox{\tt CERN-PH-TH/2006-157}
}}
{\vbox{
\ifx\answ\bigans
\vskip -5cm
\else
\vskip -4.5cm
\fi
\centerline{\hbox{Effective superpotentials for B-branes}}
\vskip 0.3cm
\centerline{\hbox{in Landau-Ginzburg models}}
}}
\vskip -.3cm
\centerline{Suresh Govindarajan$^{*,\dagger}$ and Hans Jockers$^{\dagger}$
 }
\medskip
\centerline{$^{*}${\it Institut f\"ur Theoretische Physik}}
\centerline{{\it ETH, Z\"urich, Switzerland}}
\medskip
\centerline{$^{*}${\it Department of Physics, Indian Institute of Technology
Madras}}
\centerline{\it Chennai, India}
\medskip
\centerline{$^\dagger${\it Department of Physics, Theory Division}}
\centerline{{\it CERN, Geneva, Switzerland}}

\ifx\answ\bigans
\vskip -1.0cm
\else
\vskip -.2cm
\fi

\vskip 2.5cm 
\centerline{\bf Abstract}

We compute the partition function for the topological Landau-Ginzburg
B-model on the disk. This is done by treating the worldsheet
superpotential perturbatively. We argue that this partition function as
a function of bulk and boundary perturbations may be identified with
the effective $D$-brane superpotential in the target spacetime. 
We point out the relationship of this approach to matrix factorizations.
Using these methods, we prove a conjecture for the effective superpotential
of Herbst, Lazaroiu and Lerche for the $A$-type minimal
models. We also consider the Landau-Ginzburg theory of the cubic torus 
where we show that the effective superpotential, given by the partition function, 
is consistent with the one obtained by summing up disk instantons in the 
mirror A-model. This is done by explicitly constructing the open-string 
mirror map. 

\bigskip 


\noindent

\Date{\sl {August, 2006}}

\newsec{Introduction}
Since their discovery, $D$-branes have played an important r\^ole in
many branches of string theory.  On the one hand $D$-branes appear as
ingredients in semi-realistic string model building in both particle
physics and more recently also in cosmology. On the other hand
$D$-branes have shed light on the structure of string theory from
various angles. For instance they have been crucial in revealing string
dualities and have given us some insights into non-perturbative aspects
of string theory.

Although there has been much progress over the years we still have a
rather limited understanding of $D$-branes. Just to name one example, in
the context of Calabi-Yau compactifications many aspects of $D$-branes
are only explored at the large volume point in the K\"ahler moduli
space. In this regime quantum corrections are suppressed and a
description in terms of classical geometry is applicable, whereas very
little is known at generic points in the K\"ahler moduli space. However
there exists another special point in the non-geometric region of the
K\"ahler moduli space in terms of Landau-Ginzburg orbifolds. These
theories flow in the infrared to conformal field theories (CFTs), and
for particular subclasses of Landau-Ginzburg orbifolds (which usually
correspond to special points in the complex structure moduli space)
these CFTs at the IR fixed point are known to be given by appropriate
Gepner models \refs{\GepnerVZ,\GepnerQI}.  The physics of $D$-branes in
the context of Gepner models has been investigated extensively, whereas
only recently progress has been made in studying $D$-branes in the
broader context of Landau-Ginzburg theories
\refs{\HoriCK,\KapustinBI,\BrunnerDC,\KapustinBI,\KapustinRC,\LazaroiuZI,\HoriJA}.

Already the two-dimensional bulk Landau-Ginzburg models with $(2,2)$
supersymmetry have proven useful in providing a computational framework
for CFTs and their perturbations by relevant as well as marginal
operators.  They allow us to use free-field theory methods to extract
information at the CFT end. Examples include the computation of the
central charge, the Ramond characters of minimal models and an off-shell
description of the superconformal algebra \WittenJG. 

The inclusion of $D$-branes in Landau-Ginzburg models leads one to
consider Landau-Ginzburg models with boundaries. These theories are
richer in content due to the reduced amount of supersymmetry, however,
they are also more difficult. In order to deal with this increased
complexity new tools such as the boundary linear $\sigma$-models
\refs{\GovindarajanJS,\GovindarajanKR,\HellermanBU,\HoriCK}, and matrix
factorizations have been developed
\refs{\KapustinBI,\BrunnerDC,\KapustinBI,\KapustinRC,\LazaroiuZI,\HoriJA}. 

A further understanding of the successes and limitations of
Landau-Ginzburg models is obtained from studying their topological
twisted versions. Using this approach one describes only a subsector of
the original physical string theory. This, however, also becomes a
virtue as this subsector is decoupled from all K\"ahler moduli, and as a
consequence the topological Landau-Ginzburg models describe quantities,
which have an invariant meaning in the whole K\"ahler moduli space.  In
the low-energy effective description of the underlying physical string
theory these invariants can often be identified with holomorphic
quantities protected by spacetime 
supersymmetry \refs{\BershadskyCX,\AntoniadisPM,\OoguriBV,\VafaWI,\OoguriTT}. 
Prominent
examples are prepotentials in $N=2$ theories \refs{\BershadskyCX,\AntoniadisPM} 
and gauge kinetic coupling functions and effective superpotentials in $N=1$ 
theories \refs{\BershadskyCX,\OoguriBV,\VafaWI,\DijkgraafFC,\DijkgraafVW,\OoguriTT}.

The purpose of this paper is to present a technique to compute effective
$D$-brane superpotentials by taking advantage of the computational
framework provided by (topological) Landau-Ginzburg theories. One way to
compute effective $D$-brane superpotentials in Landau-Ginzburg theories
is to study obstructions to matrix factorizations
\refs{\SiqvelandAA,\AshokXQ,\HoriJA,\KnappRD}. Then these obstructions can be
integrated to an effective $D$-brane superpotential. Although there is a
recursive algorithm in doing so \refs{\SiqvelandBB,\SiqvelandCC,\KnappRD}, 
for more involved examples the
procedure can become cumbersome. Here we provide for an alternative
approach by computing directly the effective $D$-brane superpotential
perturbatively. The key idea is to view the Landau-Ginzburg model as a
free theory, where the Landau-Ginzburg superpotential is treated
as mere perturbation \DouglasFR. From this perspective the effective
$D$-brane superpotential is simply computed by summing appropriate
Feynman diagrams of the free theory. This technique is somewhat
orthogonal to the methods discussed in the context of matrix
factorizations. However, ultimately both approaches are equivalent
\refs{\EzhuthachanJR,\EzhuthachanGU}, as we will also anticipate here.

We perform our explicit computations for the Landau-Ginzburg models of
the $A$-type minimal models and for the Landau-Ginzburg model associated
to the two-dimensional torus.  However, most of the presented analysis
is much more general and applies also for Landau-Ginzburg models of
Calabi-Yau spaces as has already been demonstrated in ref.~\DouglasFR.

The outline of the paper is as follows. In section~2 we review
Landau-Ginzburg theories without boundaries, which are capable to
describe the closed string sector. Besides fixing the notation and
introducing our two prime examples, the Landau-Ginzburg theory of the
$A$-type minimal model and of the torus, we already emphasis certain
aspects which become important later in the context of the perturbative
treatment of the Landau-Ginzburg superpotential. 

The addition of boundaries to $(2,2)$ supersymmetric Landau-Ginzburg models
is discussed in section~3. Here we also show the relation between matrix
factorizations and the boundary conditions which we impose for our
computation. Finally, we come back to our two examples for which we
describe the $D$-brane configurations considered in the forthcoming
analysis.

In section~4 we argue that the topological disk partition function of
Landau-Ginzburg models computes effective $D$-brane superpotentials.
This is demonstrated by explicitly computing the topological disk
partition function for the $A$-type minimal model at level $k$. We find
precise agreement with the effective $D$-brane superpotentials computed
for low levels of $k$ in refs.~\refs{\HerbstJP,\HerbstZM}. Our general
result confirms also the conjecture of refs.~\refs{\HerbstJP,\HerbstZM}
for the general structure of the effective superpotential in the context
of the $A$-minimal model for higher values of $k$.

In section~5 we apply the perturbative computation to the torus. This
example is much more involved due to the fact that the $D$-brane
spectrum contains now a marginal operator. As a consequence the
effective $D$-brane superpotential becomes an infinite series in terms
of the modulus associated to this marginal open-string operator. We
determine general features of the $D$-brane superpotential and we
develop the tools to explicitly compute the first few terms in this
infinite series.

In section~6 we use for the first time the modular properties of the
torus in order to gain further insight into the structure of the
effective $D$-brane superpotential for the torus example. This allows us
to confirm certain symmetry properties of the effective $D$-brane
superpotential already anticipated in the previous section.

For the two-dimensional torus there is a mirror description in the
topological A-model on the mirror torus. Thus in section~7 we map 
$D$-brane configurations of the torus in the B-model to the mirror A-model
along the lines of ref.~\PolishchukDB. Since the torus geometry is
simple enough one is able to derive the A-model $D$-brane effective
superpotential, which appears as a sum of disk instantons
\refs{\PolishchukDB,\CremadesQJ,\BrunnerMT,\HerbstNN}. By comparing the
superpotentials on both sides, we are able to construct the open-string
mirror map. Since the mirror map fulfills a set of over-constrained
equations, we obtain a highly non-trivial check on our method of
computing $D$-brane superpotentials in the topological B-model. We end
this section by some speculations on non-holomorphic terms in the
context of the topological B-model. These terms have a natural origin in
the topological A-model.

In section~8 we present our conclusions and in four appendices we
collect some technical details of various sections in the main text.

\newsec{Landau-Ginzburg Models}

In order to set the stage for the forthcoming analysis, we review the
$(2,2)$ supersymmetric Landau-Ginzburg models for the two-dimensional
worldsheet, $\Sigma$. In choosing a worldsheet, $\Sigma$, without
boundaries we describe the bulk theory or in other words the
closed-string sector of these models. A thorough introduction as well as
the detailed notational conventions used in this work can be found in
ref.~\WittenJG (also see Appendix A). 

The two-dimensional Landau-Ginzburg models with $(2,2)$ supersymmetry 
are constructed from chiral superfields, $\Phi$, which satisfy
the chirality constraint
\eqn\chirality{
\bar{D}_\alpha \Phi =0 \ , \quad \alpha=+,- \ .
}
Here $\bar{D}_\alpha$ denotes the $(2,2)$ superspace derivative. The
component fields of the chiral superfield, $\Phi$, are comprised of the
complex boson, $\phi$, the fermionic fields, $\xi$ and $\tau$, and the
complex auxiliary field, $F$.

In superspace notation the most general renormalizable action on the 
worldsheet, $\Sigma$, with several superfields, $\Phi^i$, is given by
\eqn\genLGaction{
      S_{\rm bulk} = \int_\Sigma d^2x\int d^4\theta \ K(\Phi,\bar{\Phi})
                     - \left( \lambda \ \int_\Sigma d^2x\int d\theta^+d\theta^-
                          \ W(\Phi) + {\rm h.c.} \right) \ , 
}
where $x$ are the even coordinates of the worldsheet, $\Sigma$, and
$(\theta^\pm,\bar\theta^\pm)$ are the odd coordinates of the $(2,2)$
superspace. Furthermore, the function, $K$, is the K\"ahler potential
and the holomorphic function, $W$, is the superpotential. We have also
introduced a (formal) complex coupling constant, $\lambda$, multiplying
the superpotential term.

In the topological B-model, which is the main focus of this work, two of
the four supercharges are twisted to scalar operators, $\bar Q_\pm$, by
the B-type twisting.  Hence these scalar operators are globally defined
on any closed worldsheet, $\Sigma$, and therefore they become the BRST
operators, $\bar Q_\pm$, of the topological B-model \WittenZZ.  In the
action~\genLGaction\ one finds that the first and third terms are BRST
exact. Hence, the partition function of the Landau-Ginzburg model is
expected to depend holomorphically on the coupling constants that appear
in the superpotential, $W$, and to be independent of the specific choice
of K\"ahler potential, $K$. For our computations we will make the simple
choice $K=\sum\Phi^i \bar{\Phi}_i$. We will further assume that the
superpotential is quasi-homogeneous, \ie there exist rational numbers,
$\alpha_i$, for every chiral field $\Phi^i$ such that for any
$\lambda\in {\Bbb C}^*$
\eqn\Wscaling{W(\lambda^{\alpha_i/2} \Phi^i) = \lambda\ W(\Phi^i)\ . }
The quasi-homogeneity of the superpotential ensures that we can identify
and track the left- and right-moving $R$-symmetries away from the IR
fixed point of the Landau-Ginzburg model.\foot{Since our main interest
is on worldsheets with boundaries, we will focus on the unbroken
$R$-symmetry which is the sum of the left- and right-moving $R$-charge.
This $R$-charge is given by shifting the naive free-field charge
assignment by $\alpha_i$. Thus, the $R$-charge of $\Phi^i$ equals
$\alpha_i$.} Further the central charge, $\hat c$, of the CFT is given
by
\eqn\centralcharge{ \hat{c} = \sum_i (1-\alpha_i)\ .  }
In models with several chiral fields, $\Phi^i$, we will be interested in
Landau-Ginzburg orbifolds. These are orbifolds of the above models with
a projection onto states with integral $R$-charge (the `Gepner
projection'). For specific superpotentials, $W$, such models are known
to flow in the infrared to the CFT associated with Gepner models which
in turn correspond to special points in both the K\"ahler and the
complex structure moduli space of Calabi-Yau compactifications
\refs{\GepnerVZ,\GepnerQI,\GreeneUT,\MartinecZU,\WittenYC}.

The topological observables of the bulk theory of the B-model are in the
cohomology of the BRST operators $\bar{Q}_+$ and $\bar{Q}_-$. This
cohomology is invariant under the following scaling of the
superpotential:
\eqn\CohomScaling{
W \rightarrow \lambda\ W \ . }

The quasi-homogeneity of the superpotential implies that such a scaling
can be undone by rescaling the fields $\Phi^i\rightarrow
\lambda^{-\alpha_i/2}\ \Phi^i$. This modifies the K\"ahler potential,
$K$, which is, however, an exact piece in the topological theory. Thus,
the cohomology is independent of the scaling parameter, $\lambda$.
However, it is possible that the limit, $\lambda\rightarrow0$, may be
singular. It can also be shown by studying the localization in the
topological Landau-Ginzburg model that the parameter $\lambda$ can be
identified with the renormalization scale with $\lambda \rightarrow 0$
being the UV limit and $\lambda \rightarrow \infty$ being the 
IR limit \refs{\VafaMU,\HerbstAX}. 

The topological BRST operator
localizes on the space of zero-modes and takes the following form
\refs{\HoriCK,\HerbstAX}:
\eqn\redBRST{
\cQ\sim (\bar{Q}_+ + \bar{Q}_- )|_{\rm zero modes}\sim
\bar{\partial} + i_{\partial W} \ .
}
Here $\bar{\partial}$ is the Dolbeaut operator of the non-compact
target space, $X$, of the action~\genLGaction, 
while the operator, $i_{\partial W}$, acts upon sections of the graded
space, $\wedge^\bullet TX$.\foot{
Locally the sections of $\wedge^\bullet TX$ are obtained as 
wedge products of vector fields.}
The definition of $i_{\partial W}$ is induced by its action on vector fields 
\eqn\iDW{i_{\partial W}(v^j\partial_j)= -iv^j \partial_j W \ , }
which naturally extends to a general sections of the graded space,
$\wedge^\bullet TX$, on which the operator, $i_{\partial W}$, becomes
the odd derivation appearing in the BRST operator~\redBRST. Since
$\bar{\partial}^2 =i_{\partial W}^2 = 0$ and $\bar{\partial}i_{\partial
W} +i_{\partial W}\bar{\partial}=0$, the topological observables are
given by the double cohomology of the two differentials,
$\bar{\partial}$ and $i_{\partial W}$. This usually involves computing a
spectral sequence whose second term is $E_2 = H_{i_{\partial
W}}(H_{\bar{\partial}}(X))$.  In some situations, \eg when the space, $X$, 
is given by $X=\IC^n$ and when the superpotential, $W$, 
is a polynomial in these $n$ variables, the cohomology of $\cQ$
is equal to $E_2$, \ie $E_2=E_\infty$. In other words the dimension of 
the double cohomology is simply determined by the cohomology of 
$i_{\partial W}$ within the cohomology group $H_{\bar\partial}(X)$ \KachruPG. 
Then we can treat the Landau-Ginzburg superpotential, $W$, perturbatively
at the level cohomology and, as we will verify in the discussed examples,
also at the level of correlation functions. 

\subsec{Example 1: The $A$-type minimal model}
In this paper we will consider two examples. The first one is the
$A$-type mini\-mal model at level $k$. The Landau-Ginzburg description
consists of a single chiral superfield, $\Phi$, and the Landau-Ginzburg
superpotential
\eqn\superone{
W_{A_k} = \frac{\Phi^{k+2}}{k+2} - \sum_{j=2}^{k+2} g_j(t)\ \Phi^{k+2-j}\ .
}
The coupling constants $g_j(t)=t_j+\cdots$ parametrize relevant bulk
deformations about the conformal point and are taken to be functions of
the flat coordinates $t_2,\ldots,t_{k+2}$ \DijkgraafDJ. The conformal
point is given by $g_j=0$. It is useful to treat the coefficient,
$1/(k+2)$, of $\Phi^{k+2}$ as the coupling, $g_0$.

As a particular example consider the $k=3$ model for which the
Landau-Ginzburg superpotential takes the form
\eqn\kthree{
W_{A_3}= \frac{\Phi^5}{5}-t_2 \Phi^3-t_3 \Phi^2-\left(t_4-t_2^2\right)
\Phi -(t_5 - t_2 t_3)\ .
}
The $R$-charge assignments for the component fields of the chiral
multiplet, $\Phi$, resulting from eq.~\Wscaling\ at the conformal point
are summarized in \ltab\ARcharges.

\tabinsert\ARcharges{$R$-charges for the $A_k$-minimal model in terms of $\alpha={2\over k+2}$.}
{
\begintable
 Field      \| \hskip 3ex $\phi$\hskip 3ex \| \hskip 3ex $\tau$ \hskip 3ex
            \| \hskip 3ex $\xi$ \hskip 3ex \| \hskip 3ex $F$ \hskip 3ex     \cr
 $R$-charge \| $\alpha$ \| $\alpha-1$ \| $\alpha-1$ \| $\alpha-2$
\endtable }

\subsec{Example 2: The cubic torus}\subseclab\bulktorus

The second example is the Landau-Ginzburg orbifold that flows in the
infrared to the CFT, which describes strings on the two-dimensional
torus, $\cT$, at the Landau-Ginzburg point of the K\"ahler moduli space.
This Landau-Ginzburg model consists of three chiral superfields, $\Phi^i$, the 
cubic Landau-Ginzburg superpotential
\eqn\Wtorus{W = \sum_{i=1}^3 \left(\Phi^i\right)^3  
- 3\,a\,\Phi^1\Phi^2\Phi^3 \ , }
and the $\IZ_3$-orbifold action, $\Phi^i\rightarrow e^{2\pi i/3}\
\Phi^i$. The coupling, $a$, parametrizes the complex structure of the
cubic torus, $\cT$.  At the Fermat point, \ie for $a=0$, in the complex
structure moduli space the Landau-Ginzburg orbifold flows in the
infrared to the $1^3$ Gepner model, \ie it flows to the CFT obtained
form the tensor product of three $k=1$ $A$-type minimal models subject
to the Gepner projection \refs{\GepnerVZ,\GepnerQI}. Therefore, as
summarized in \ltab\TRcharges, the $R$-charges of the fields in the
chiral multiplets, $\Phi^i$, of the cubic torus, $\cT$, coincide with
the $R$-charges of chiral multiplet in the $A$-type minimal model at
level $k=1$.

\tabinsert\TRcharges{$R$-charges for the Landau-Ginzburg orbifold of the torus.}
{
\begintable
 Field      \| \hskip 3ex $\phi$\hskip 3ex \| \hskip 3ex $\tau$ \hskip 3ex
            \| \hskip 3ex $\xi$ \hskip 3ex \| \hskip 3ex $F$ \hskip 3ex     \cr
 $R$-charge \| $2\over 3$ \| $-{1\over 3}$ \| $-{1\over 3}$ \| $-{4\over 3}$
\endtable }

For our forthcoming computation it is convenient to rewrite the
superpotential~\Wtorus\ in terms of two independent coupling constants,
$g_0$ and $g_1$, which in turn appear in the coupling tensor, $c_{ijk}$:
\eqn\Wtorus{W = g_0\,\left(\sum_{i=1}^3 \left(\Phi^i\right)^3\right)
                + g_1\left(-3\,\Phi^1\Phi^2\Phi^3\right)
              = \sum_{i,j,k} c_{ijk} \Phi^i\Phi^j\Phi^k \ . }
The couplings, $c_{ijk}$, are symmetric in all indices and are given by
\eqn\Defc{ c_{ijk} = \cases{\hphantom{-}g_0 & for $i=j=k$ \cr
                            -{g_1\over 2}   & for $i\ne j,j\ne k,i\ne k$ \cr
                            \hphantom{-}0   & else \ .}}
Note that the original complex structure coupling, $a$, is now identified with
\eqn\Moda{ a = {g_1\over g_0} \ . }

The parameter, $a$, in the
superpotential, $W$, 
is related to the standard complex structure
modulus, $\tau$, of the torus via the relation \LercheCS
\eqn\Defa{j(\tau)=\left({3\,a\,(a^3+8)\over a^3-1}\right)^3 \ . }
Here $j(\tau)$ denotes the modular invariant $j$-function of the torus.
As in the minimal models, $\tau$ appears as the flat coordinate.
Naively, the overall scaling of $g_0$ and $g_1$ are not important since
only their ratio gives the complex structure modulus, $\tau$, of the
torus. However, to parametrize the marginal deformations of the
Landau-Ginzburg superpotential in terms of flat coordinates, it is
necessary to adjust the normalization of the Landau-Ginzburg
superpotential by an appropriate `flattening' factor in order to ensure
the vanishing of the Gauss-Manin connection \LercheWM.  For the cubic
superpotential~\Wtorus\ the appropriate normalization is given by
$g_0=\qf^{-1}$, $g_1=\qf^{-1} a$, where the `flattening' factor is
\LercheWM
\eqn\Defqf{\qf(\tau)=\sqrt{1-a^3(\tau)\over 3\,a'(\tau)}
                    ={1\over 3\sqrt{2\pi i}}\,{\eta(\tau)\over\eta^3(3\tau)} \ . }
We will find later that this choice leads to simplifications in the 
open-string sector as well.

\newsec{Landau-Ginzburg models with boundary}

In order to describe branes in the Landau-Ginzburg theory~\genLGaction,
we consider worldsheets, $\Sigma$, with boundaries, $\partial\Sigma$. As
we will see the worldsheets relevant for our analysis have a single
boundary. In other words $\Sigma$ has the topology of a disk, which we
can map to the complex upper-half plane with coordinates $(x,y)$ with
$x\in (-\infty,+\infty)$ and $y\in [0,+\infty)$. Thus, the single
boundary, $\partial\Sigma$, is just the line $y=0$. Here we are
interested in B-branes and therefore the boundary conditions on the
worldsheet, $\Sigma$, must be compatible with the B-twist. From a
superspace point of view this means that the two-dimensional $(2,2)$
superspace of the bulk theory reduces to the one-dimensional boundary
superspace (with two Grassmann coordinates
$\theta=\frac{\theta^--\theta^+}{\sqrt2}$ and
$\bar{\theta}=\frac{\bar{\theta}^--\bar{\theta}^+}{\sqrt2}$).  Then a
chiral superfield, $\Phi$, restricted to the boundary, $\partial\Sigma$,
becomes a boundary chiral superfield, $\Phi_\partial$, which obeys the
boundary chirality constraint
\eqn\BdryConst{\bar D\Phi_\partial=0 \quad{\rm with}\quad
               \bar D={\partial\over\partial\bar\theta}-i\theta\partial_x \ . }
This constraint implies that at the boundary the boson, $\phi$, and the
fermion, $\tau$, are the non-vanishing
components of the restricted multiplets, $\Phi_\partial$, which have the
expansion
\eqn\chiralbdry{\Phi_\partial=\phi+\sqrt{2}\theta\tau-i\theta\bar\theta
                \partial_x\phi \ . }

Varying the Landau-Ginzburg action~\genLGaction\ with respect to the
supercharge to be preserved at the boundary, one finds, however, that
the bulk superpotential, $W$, generates a non-vanishing boundary term,
which is often called the Warner term \WarnerAY
\eqn\VarLG{
   \delta_{\epsilon} S \sim
      \lambda\,\int_{\partial\Sigma} dx\,\bar\epsilon\
         \sum_i \partial_i W(\phi)\ \tau^i\ +\ {\rm h.c.} \  . }
Here $\epsilon,\bar\epsilon$ are the infinitesimal fermionic parameters
of the supersymmetry variation. In order to preserve the B-type supersymmetry 
there are two possibilities to proceed. 

\noindent
(i) Impose boundary conditions on the bulk superfields, $\Phi$, such
that the Warner term vanishes. This corresponds to choosing boundary
conditions that imply $W=0$ on the boundary \refs{\WarnerAY,\GovindarajanJS}.

\noindent
(ii) Add boundary superfields with a boundary action whose supersymmetry
variation cancels the Warner term \refs{\GovindarajanKR,\HellermanBU}.
This approach naturally leads to the Kontsevich's description of
B-branes in terms of matrix factorizations of the superpotential
\refs{\KapustinBI,\BrunnerDC}.

Recent work has provided evidence for the equivalence of the first
possibility to a subclass of matrix factorizations
\refs{\EzhuthachanJR,\EzhuthachanGU}.  This subclass can be represented
by boundary superfields and in the two examples we consider, generate
all other matrix factorizations by tachyon condensation
\refs{\HerbstZM,\GovindarajanIM}. 

\subsec{Matrix Factorizations}

We will now briefly review matrix factorizations from the viewpoint of
adding boundary superfields.  Given a factorization of the
superpotential, $W$, of the form
\eqn\Factors{W(\phi)=\sum_a J^a(\phi)\, E_a(\phi) \ , \quad a=1,\ldots,k \ ,}
we construct a boundary action, which compensates the Warner
term~\VarLG. This is achieved in terms of $k$ boundary fermionic
superfields, $\Pi_a$, which obey the superspace constraint
\eqn\BdryFermions{\bar D\Pi_a = \lambda\,E_a(\Phi_\partial) \ . }
The fermionic boundary multiplets, $\Pi_a$, are comprised of boundary
fermions, $\pi_a$, and bosonic auxiliary fields, $\ell_a$. The kinetic
terms for these fields are given by:
\eqn\BdryKin{S_{\rm kin} = \int_{\partial\Sigma} dx\int d^2\theta\, \sum_a \bar\Pi^a\Pi_a \ . }
In component fields the auxiliary fields, $\ell_a$, appear only
algebraically, whereas the kinetic terms of the boundary fermions,
$\pi_a$, give rise to the one-dimensional Dirac equation. Finally, we
also add a boundary superpotential to the boundary action
\eqn\BdryW{S_J = \int_{\partial\Sigma} dx \int d\theta\,\sum_a\Pi_a 
J^a(\Phi_\partial) + {\rm h.c.} \ . }
Due to the modified chirality constraint~\BdryFermions\ this boundary
superpotential is not supersymmetric by itself. It is straight forward
to check that due to the factorization~\Factors\ the supersymmetry
variation of the boundary superpotential cancels the Warner term~\VarLG.
Hence the boundary superpotential, $J$, together with the
constraint~\BdryFermions\ imposed on the superfields, $\Pi_a$, are the
important ingredients, which are needed to preserve the B-type
supersymmetry in the Landau-Ginzburg theory with boundaries.

Since the bulk action~\genLGaction\ together with the boundary
action~\BdryKin\ and \BdryW\ does now preserve the B-type supercharge, we
can perform the B-twist and obtain the topological B-model with
boundaries. To the bulk BRST operator~\redBRST\ we now need to add the
boundary BRST operator, $\cQ_{\partial}$, which acts on the boundary
fields, $\pi_a$ and $\bar\pi^a$, as\foot{Here we have eliminated the
auxiliary fields, $\ell_a$ and $\bar\ell^a$.}
\eqn\BdryBRST{\cQ_{\partial}\pi_a = \lambda\,E_a(\phi) \ , \quad
             \cQ_{\partial}\bar\pi^a = J^a(\phi) \ . }

In the boundary action, $S_{\rm kin}$, the fermionic fields, $i
\bar\pi^a$, are conjugate to the fermionic fields, $\pi_a$. Hence, in
the canonically quantized boundary theory the boundary fermions obey the
canonical anti-commutation relation
\eqn\CanACom{\{\pi_a,\bar\pi^b\}=\delta_a^b \ . }
Therefore the boundary BRST operator, $\cQ_{\partial}$, can be written as
\eqn\BRSTBdryOperator{\cQ_{\partial}^\lambda = J + \lambda\,E \ , }
in terms of the operators, $J$ and $E$, defined by
\eqn\OpJE{J(\phi,\pi)=\sum_a J^a(\phi) \pi_a \ , \quad
          E(\phi,\bar\pi)=\sum_a E_a(\phi) \bar\pi^a \ . }

This structure of the boundary BRST operator is now directly related to
the description of B-branes in terms of matrix factorizations.
Namely, choosing a matrix representation for the Clifford
algebra~\BdryBRST\ generated by the boundary fermions, $\pi_a$ and
$\bar\pi^a$, we obtain a $2^k\times 2^k$-matrix representation of the
boundary BRST operator, $\cQ_\partial$. Furthermore, in the basis, in
which the chirality matrix, $\gamma$, is diagonal
\eqn\CMat{\gamma=
      \pmatrix{{\bf 1}_{2^{k-1}\times 2^{k-1}} & 0 \cr
               0 & -{\bf 1}_{2^{k-1}\times 2^{k-1}} } \ , }
the boundary BRST operator, $\cQ_\partial$, can be expressed in terms of
two $2^{k-1}\times 2^{k-1}$ matrices, $G^\lambda(\phi)$ and $F^\lambda(\phi)$:
\eqn\BRSTmatrix{\cQ_\partial^\lambda = \pmatrix{0 & G^\lambda(\phi) \cr
                                        F^\lambda(\phi) & 0 } \ . }
From the definition~\BRSTBdryOperator\ of the boundary BRST operator we
learn that the matrix, $\cQ_\partial$, squares to the superpotential,
$\lambda W$. This is equivalent to imposing
\eqn\FGmatrix{
F^\lambda(\phi)\cdot G^\lambda(\phi)= G^\lambda(\phi)\cdot F^\lambda(\phi)
=\lambda\,W(\phi)\ {\bf 1}_{2^{k-1}\times 2^{k-1}}\ ,
}
which is a matrix factorization in the Kontsevich sense \KapustinBI.
Hence such matrix factorizations~\BRSTmatrix\ of the superpotential
yield an equivalent description of the boundary BRST
operator~\BRSTBdryOperator.\foot{At first glance it seems that there are
only $2^{k-1}\times 2^{k-1}$-matrix factorizations. However, these are
often equivalent to lower-dimensional matrix factorizations by
condensing trivial brane anti-brane pairs
\refs{\HerbstZM,\GovindarajanIM}.}

The cohomology of the boundary BRST operator~\BdryBRST\ gives rise to
the open-string operators in the topological theory, \ie these operators
are non-trivial cohomology elements with respect to the differential,
${\cal D}^\lambda$, which acts upon an operator, $\Psi$, as
\eqn\Dop{{\cal D}^\lambda(\Psi) = [ \cQ_\partial^\lambda, \Psi ]_\pm \ . }
The commutator applies for bosonic operators whereas the anti-commutator
is taken for fermionic operators.

When each of the $J^a$ and $E_a$ are quasi-homogeneous, the boundary
fermionic multiplets can be assigned $R$-charges. Further, the
cohomology of $\cD^\lambda$ can be shown to be independent of $\lambda$.
However, the limit $\lambda\rightarrow 0$ can be singular and the
cohomology at $\lambda=0$ need not agree with the cohomology for
non-zero $\lambda$. To study this aspect, we observe that the boundary
operators, $J$ and $E$, obey
$[\{J,J\},\Psi]_\pm=[\{E,E\},\Psi]_\pm=[\{J,E\},\Psi]_\pm=0$. Hence the
individual operators, $J$ and $E$, give rise to two commuting cohomology
differentials, $[J,\,\cdot\,]_\pm$ and $[E,\,\,\cdot\,]_\pm$.  Therefore
in computing the cohomology at $\lambda=0$, that is to say the
cohomology of the differential, ${\cal D}^{\lambda=0}\equiv
[J,\,\cdot\,]_\pm$, we get a first approximation for the cohomology with
respect to ${\cal D}^{\lambda\ne 0}$. However, in correcting the
cohomology elements perturbatively in $\lambda$, some cohomology
elements of ${\cal D}^{\lambda=0}$ might drop out at a finite order in
$\lambda$ because at this order they cannot be completed perturbatively
to a cohomology element of the full BRST operator, ${\cal D}^{\lambda\ne
0}$. This process of recursively completing the cohomology elements
corresponds to evaluation the spectral sequence of the double complex
associated to the commuting differentials, $[J,\,\cdot\,]_\pm$ and
$[E,\,\,\cdot\,]_\pm$. In the two examples that we consider in this
paper, it turns out that the limit $\lambda \rightarrow 0$ is {\it not}
singular.

We will now discuss the boundary conditions and the corresponding
matrix factorizations in the two examples of interest.

\subsec{Example 1: The $A$-type minimal model}

The only possible boundary condition in the Landau-Ginzburg model~\superone\ 
that flows in the infrared to the $A$-type minimal model is 
the Dirichlet boundary condition \GovindarajanJS, 
\eqn\bcminimal{
\Phi_\partial =c\ .
}
The constant, $c$, is equal to any root of $W_{A_k}$, and therefore we
obtain $W_{A_k}|_{\partial\Sigma}\equiv 0$. Furthermore, as discussed in
Appendix~A, the Dirichlet boundary condition implies that the remaining
components of the chiral superfield, $\Phi$, (and its conjugate,
$\bar\Phi$) form in the absence of the superpotential a fermionic
boundary chiral superfield, $\bar\Xi$, with the superfield expansion
\eqn\bsuper{
\bar{\Xi} = \bar{\xi} - \sqrt2 \theta i \partial_y\bar{\phi} 
-i\theta\bar{\theta} \partial_x \bar{\xi}\ .
}

The boundary chiral superfield, $\bar\Xi$, gives rise to interactions at
the boundary, $\partial\Sigma$, of the worldsheet, $\Sigma$. The only
possible boundary interaction, which is relevant in the topological
theory, is given by
\eqn\bint{
S_\partial = X\ \int_{\partial\Sigma} dx \int d\theta\ \bar{\Xi}\ +\ {\rm h.c.} \ . }
We have introduced a boundary coupling constant, $X$, which we will eventually 
promote to a coupling matrix so as to include Chan-Paton factors.

The boundary condition~\bcminimal\ is equivalent to the rank-one matrix
factorization with $J=\phi$ and $E=\phi^{k+1}/(k+2)$, which again is
equivalent to the $L=0$ boundary state in the corresponding boundary CFT
\BrunnerDC. We observe that all three descriptions have precisely one
boundary deformation with identical $R$-charges. This is a simple check
on the equivalence of those three formulations. A more intuitive way to
understand the connection to matrix factorizations is to see that
$\bar{\xi}$ effectively plays the r\^ole of the boundary fermion, $\pi$,
and the boundary condition~\bcminimal\ becomes the low-energy condition
$J=0$. As we will discuss further in our second example, in order to match
the boundary deformations in these different formulations, a certain spectral
sequence associated to the double complex of the boundary operators,
$J$ and $E$, needs to collapse. This is similar to the situation arising 
from the double complex associated to the bulk BRST operator~\redBRST.

It is known that the matrix factorizations for the $L>0$ boundary states
in the CFT can be obtained by tachyon condensation of a suitable number
of $L=0$ matrix factorizations \HerbstZM.  In our approach several
boundary components can be added by using Chan-Paton indices to
distinguish the boundaries. The various coupling constants now become
coupling matrices carrying Chan-Paton indices. For example, in addition
to imposing Dirichlet boundary conditions the $L=1$ matrix factorization
is represented by enhancing the coupling constant, $X$, to the $2\times
2$ coupling matrix, $\cmat{X}$. Then the off-diagonal entries of
$\cmat{X}$ representing `tachyons' that mediate the formation of the
$L=1$ bound state.

\subsec{Example 2: The `Long' branes on the cubic torus}

\tabinsert\LongCohom{Fermionic and bosonic operators in the
cohomology of the differential, ${\cal D}^{\lambda=0}$.
All these operators extend to the cohomology of ${\cal D}^{\lambda\ne 0}$,
where the cohomology elements in the columns are related by Serre duality.}
{
\begintable
 Fermionic operators   \| \hskip 3ex $\pi_1, \pi_2, \pi_3$\hskip 3ex
                       \| \hskip 3ex $\pi_1\pi_2\pi_3$ \hskip 3ex  \cr
 Bosonic operators     \| \hskip 3ex $\pi_1\pi_2, \pi_1\pi_3, \pi_2\pi_3$ \hskip 3ex
                       \| \hskip 3ex ${\bf 1}$ \hskip 3ex
\endtable }

On the cubic torus we focus in this work on the brane configuration which
is associated to a $4\times 4$-matrix factorization generated by
three boundary fermions, $\pi_i$:
\eqn\BRSTfour{\cQ_\partial^\lambda = \phi^i \pi_i + \lambda\,\partial_i W(\phi)\,\bar\pi^i \ ,
              \quad i=1,2,3 \ . }
This matrix factorization describes the `long' branes, $L_a$, in the
terminology of refs.~\refs{\BrunnerMT,\GovindarajanIM}.\foot{One obtains
three `long' branes, $L_a$, since one really considers equivariant
matrix factorizations, which introduces the additional $\IZ_3$-valued
label $a$ \refs{\AshokZB,\GovindarajanIM}.} The cohomology at
$\lambda=0$ is readily computed and is summarized in \ltab\LongCohom.
Furthermore for the matrix factorization~\BRSTfour\ it is easy to check
that all the cohomology elements for $\lambda=0$ can be recursively
completed to cohomology elements for finite values of $\lambda$, and
hence the dimension of the cohomology group is not changed by the
Landau-Ginzburg superpotential. This simplification need not occur for a
generic matrix factorization, \eg for the `short' branes of
refs.~\refs{\BrunnerMT,\GovindarajanIM}\ the boundary cohomology of the
free theory is larger than the cohomology in the presence of the
Landau-Ginzburg superpotential. Here we concentrate on the `long'
branes, $L_a$, where this subtlety does not play a r\^ole, but we come
back to the general situation elsewhere.

\figinsert\torusrels{The relationship between the different constructions
of the `long' branes on the torus, $\cT$, at the Gepner point in the K\"ahler
moduli space.
}{1.3in}{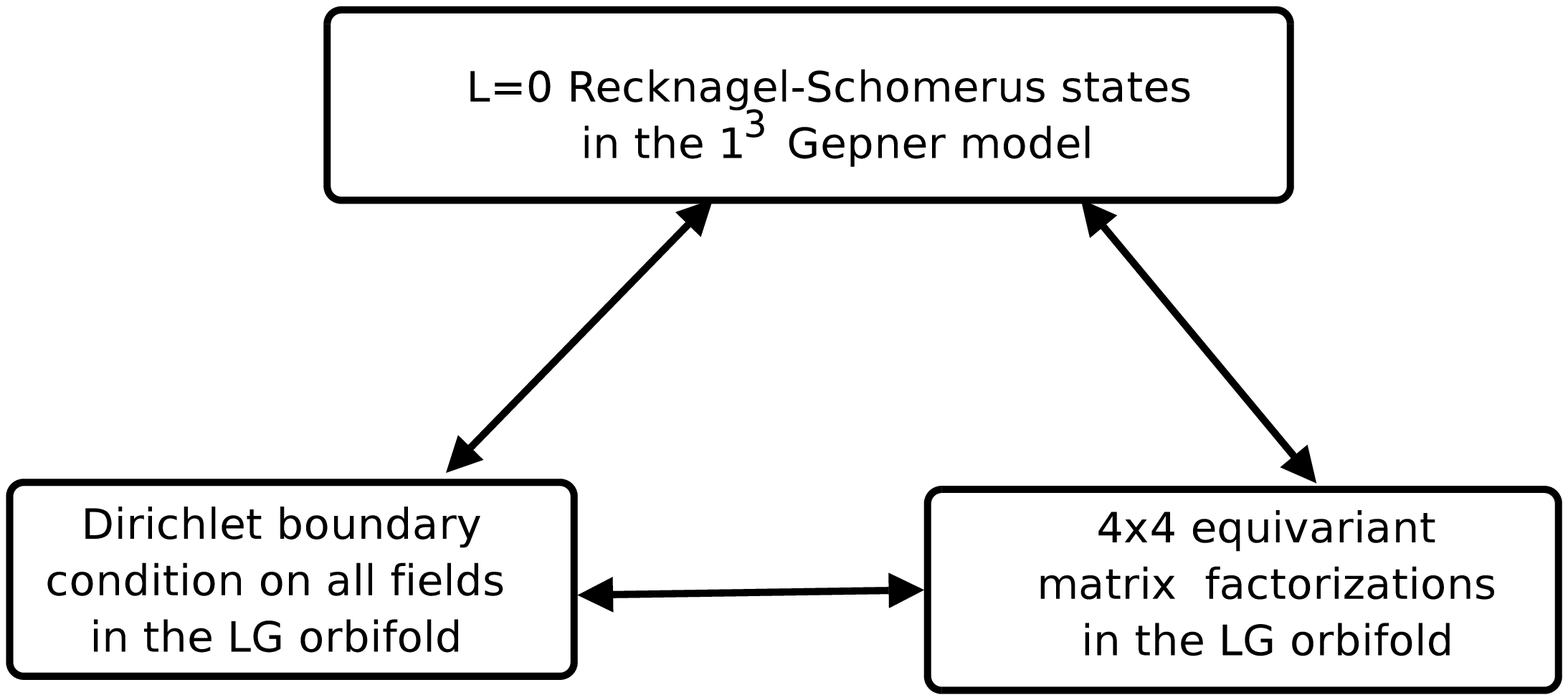}

Similarly to the equivalent descriptions of branes in the $A$-type
minimal model there are also different formulations for the `long'
branes, $L_a$, on the cubic torus as summarized in \lfig\torusrels. 
Here we mainly describe the `long' branes by imposing Dirichlet boundary
conditions.  Therefore, in order to close the circle we sketch the
relationship of the $4\times 4$-matrix factorization~\BRSTfour\ of the
torus to the brane description obtained by imposing Dirichlet boundary
conditions on all three chiral bulk multiplets, $\Phi^i$. As discussed
in the previous subsection and in Appendix~A, Dirichlet boundary
conditions correspond to $\phi^i=0$ and $\tau^i=0$, and on the boundary
they reduce the bulk chiral multiplets, $\Phi^i$, to a chiral fermionic
boundary multiplets, $\bar\Xi_i=(\bar\xi_i,\partial_y \bar\phi_i)$
\DouglasFR. 

With these boundary fermions, we can now construct boundary operators.
In particular there are two kinds of fermionic operators, namely
$\psi=\cmat{X}^i \bar{\xi}_i$ and $\Omega=\cmat{U}\epsilon^{ijk}
\bar{\xi}_i \bar{\xi}_j \bar{\xi}_k$, which match the results from
matrix factorization summarized in \ltab\LongCohom. Furthermore, from
the scaling~\Wscaling\ of the Landau-Ginzburg superpotential we know
that the chiral bulk superfields, $\Phi^i$, have $R$-charge $+{2\over3}$ 
(\cf \ltab\TRcharges).  This allows us to infer that the operators, $\psi$
and $\Omega$, have $R$-charges $+{1\over 3}$ and $+1$, and therefore they
correspond to relevant operators and a marginal operator, respectively.

In order to describe interactions among several branes the couplings,
$\cmat{X}^i$ and $\cmat{U}$, are not just taken to be scalars, but
instead they are enhanced to Chan-Paton matrices. In the absence of the
superpotential the Landau-Ginzburg orbifold of the torus degenerates to
the $\IC^3/\IZ_3$ orbifold, which has three fractional $D0$-branes
\EzhuthachanJR. Here these three branes are distinguished by assigning
to the boundary three different Chan-Paton labels, and the matrix-valued
couplings, $\cmat{X}^i$ and $\cmat{U}$, become
\eqn\chanpaton{
     \cmat{X}^i=\pmatrix{0 & x^i_{12} & 0 \cr 0 & 0 & x^i_{23} \cr x^i_{31} & 0 & 0 } \quad,\quad
     \cmat{U}  =\pmatrix{u_1 & 0 & 0 \cr 0 & u_2 & 0 \cr 0 & 0 & u_3} \ . }
The off-diagonal nature of the Chan-Paton matrices, $\cmat{X}^i$, implies that
the operators, $\psi$, are boundary condition changing operators
while the diagonal nature of the matrix, $\cmat{U}$, implies that the operator,
$\Omega$, is a boundary condition preserving operator. The boundary
operators with their Chan-Paton matrices reflect the quiver diagram
associated to the three fractional $D0$-branes of the $\IC^3/\IZ_3$
orbifold \DouglasQW. In the matrix factorization picture, the 
orbifolding of the Landau-Ginzburg model
implies that we must deal with equivariant matrix factorizations and
the Chan-Paton indices are the equivariant 
labels \refs{\AshokZB,\WalcherTX,\GovindarajanIM}. 

The boundary interaction that is relevant to the topological theory on
the cubic torus is given by
\eqn\bintb{
S_\partial = \mu\ \cmat{X}^i\int_{\partial\Sigma} dx \int d\theta\ \bar{\Xi}_i 
           + \nu\ \cmat{U}  \int_{\partial\Sigma} dx \int d\theta\ \epsilon^{ijk} 
              \bar{\Xi}_i \bar{\Xi}_j \bar{\Xi}_k \ +\ {\rm h.c.} \ ,
}
where $\mu$ and $\nu$ are two constant parameters that we introduce for future
convenience.

\newsec{The topological partition function and the effective superpotential}

The main focus of this work is to compute in the context of
Landau-Ginzburg models the topological partition function, $Z^{\rm
top}_{\rm disk}$, for worldsheets with the topology of a disk. This
partition function is holomorphic in both the bulk and boundary
couplings, and it is directly related to the worldvolume superpotential,
$\cW$, of the physical open-superstring theory. In this section we
introduce the topological partition disk function, $Z^{\rm top}_{\rm
disk}$, and show its connection to the effective $D$-brane
superpotential, $\cW$. We conclude this section by illustrating these
concepts with the $A$-type minimal model.

\subsec{Interpreting the topological partition function}

We will now argue that the topological partition function on the disk,
$Z^{\rm top}_{\rm disk}$, computes the effective superpotential, $\cW$,
on the worldvolume of the corresponding brane configuration. Based on
the results of Shatashvili and Witten \refs{\WittenQY,\ShatashviliKK},
it was conjectured in ref.~\KutasovAQ\ and proved in ref.~\NiarchosSI\
that the tree-level effective action of the open-superstring field
theory is given by the superstring disk partition function.  
Thus, in the topological theory,
which describes the holomorphic subsector of the physical theory, it is
natural to make the identification of the topological partition
function of the Landau-Ginzburg model, 
$Z^{\rm top}_{\rm disk}$, with the holomorphic effective
$D$-brane superpotential, $\cW$ \BershadskyCX: 
\eqn\ZW{Z^{\rm
top}_{\rm disk}(t,u)=\cW(t,u)\ . 
} 
Here the the complex parameters, $t$ and $u$, symbolically represent the
bulk and boundary couplings, respectively. 

Obstructions to $D$-brane deformations give rise to yet another relation
to the topological partition function. In the context of branes given as
matrix factorizations in Landau-Ginzburg models, obstructions in
deforming matrix factorization can be encoded in a superpotential
\refs{\AshokXQ,\HoriJA,\KnappRD}, which we denote by $\cW_{\rm def}$.
This means that matrix factorizations exist only on the sublocus in the
space of bulk and boundary couplings, $t$ and $u$, where $d\cW_{\rm
def}(t,u) =0$. For the two examples that we study, we will find that up
to field redefinitions
\eqn\ZWdef{Z^{\rm top}_{\rm disk}(t,u)=\cW_{\rm def}(t,u)\ , }
where $u$ parametrizes deformations of the boundary BRST operator,
$\cQ_\partial$, by fermionic operators, whereas $t$ captures the
(bulk) deformations of the Landau-Ginzburg superpotential, $W$. This approach
to superpotentials provides an interpretation that works also for the
non-geometric examples such as the $A$-type minimal model, and it is
also in agreement with the suggestion of
refs.~\refs{\BrunnerJQ,\KachruIH}\ that the lifting of moduli arising
from vector bundles is encoded in effective superpotentials.

Finally, $Z^{\rm top}_{\rm disk}$ has a third interpretation as the
generating function of (symmetrized) correlation functions
\refs{\HerbstJP,\HerbstZM,\AspinwallBS}. This follows formally from the
path-integral representation of the topological disk partition function
because taking derivatives with respect to the coupling constants and
then setting the couplings to zero is equivalent to computing
(symmetrized) correlation functions.

\subsec{A perturbative computation of the effective superpotential}
The main advantage of the Landau-Ginzburg model is that some of the
computations can be reduced to those involving free fields. In other
words, one can treat the Landau-Ginzburg superpotential, $W$,
perturbatively. This means that we go to the limit $\lambda=0$, in which
the Landau-Ginzburg superpotential, $W$, vanishes, and then we perform a
perturbative expansion in the formal coupling constant, $\lambda$. In
doing so we take advantage of the free-field formulation and in
particular following ref.~\DouglasFR\ we also use the bulk-boundary
propagators of free-fields. We restrict our attention to Dirichlet
boundary conditions, which are the simplest boundary conditions in
Landau-Ginzburg models, and which, as discussed in the previous section,
are compatible with the perturbative limit, $\lambda=0$, due to the fact
that the boundary cohomology does not become singular.  Thus altogether
one has
\eqn\diskpartfnb{
\cW\equiv Z_{\rm disk}^{\rm top}=
 \sum_{n=0}^\infty \lambda^n \left\llangle \frac{(S_W)^n}{n!}\
P\left(e^{\cal S_\partial}\right)\right\rrangle \equiv \sum_{n=0}^\infty \lambda^n \cW_n\ ,
}
where $\langle\!\langle \cdots \rangle\!\rangle$ denotes the free-field
correlators on the disk, $\Sigma$, which has been mapped to the upper
half plane. Further, $S_W=\int_\Sigma d^2x \int d^2 \theta W(\Phi)$ and
$S_\partial$ represent bulk and boundary interactions. The latter appear
path-ordered, $P(\,\cdot\,)$, because, as explained in the previous
section, including Chan-Paton factors renders the boundary interactions,
$S_\partial$, matrix-valued. 

\subsec{The $A$-minimal model} \subseclab\firstcomp
We will now proceed to compute the effective $D$-brane superpotential,
$\cW$, or equivalently the topological disk partition function, $Z_{\rm
disk}^{\rm top}$, in the Landau-Ginzburg model for the $A$-minimal
model. This is in many ways the simplest family of Landau-Ginzburg
models. Naively, the topological partition function vanishes because of
fermionic zero-modes. Therefore only correlators which saturate those
zero-modes contribute to the topological partition function. In addition
we also need to gauge-fix the $PSL(2,\IR)$ invariance of the upper-half
plane. So, we compute $\frac{\partial^2 \cW}{\partial \lambda \partial
X}$, which is according to eq.~\diskpartfnb\ a two-point function
involving one bulk operator and one boundary operator. The $PSL(2,\IR)$
invariance is fixed by choosing these two operators as unintegrated
(zero-form) operators. To be specific, we place the bulk operator at
$(x_0,y_0)$ and the boundary operator at $x=+\infty$. The boundary
zero-form operator is given by $\bar\xi(+\infty)$ and thus also provides
for the required $\bar{\xi}$-zero mode. Thus, we will compute
\eqn\twopointa{
\frac{\partial^2 \cW}{\partial \lambda \partial X} =
\sum_{j=0}^{k+2}
\left\langle V_j^{(0)}(x_0,y_0)\ \bar{\xi}(+\infty) \right\rangle \ ,
}
where $V_j^{(0)}=g_{k+2-j}(t)(\phi)^j$ is the bulk zero-form operator.

In computing the free-field correlation functions, non-vanishing
correlators appear only if the total $R$-charge of all operators equals
$\hat{c}=(1-\alpha)$.  Further, with the exception of the fermionic zero
modes all the fields that appear in the operators must be contracted
with (some) other field in order to yield a non-zero answer. A simple
consideration of these two conditions shows that for this example,
correlators involving more than one bulk insertion vanish. Thus, $\cW_1$
is the only non-vanishing contribution to the effective superpotential,
$\cW$, as defined in eq.~\diskpartfnb, and we find
\eqn\WlX{
\frac{\partial^2 \cW}{\partial \lambda \partial X} =
\sum_{j=0}^{k+2}
g_{k+2-j}(t) \frac{X^{j}}{j!} \left\llangle (\phi)^j(x_0,y_0)
P\left(\frac{1}{\sqrt2}\prod_{k=1}^j\int\limits_{-\infty}^{+\infty}
dx_k\ i\partial_y\bar{\phi}(x_k)\right)  \bar{\xi}(\infty) \right\rrangle \ . }
Since we assume for now that the coupling, $X$, is a scalar, path-ordering 
of the boundary operators is not necessary.

The next task is to explicitly evaluate the correlator~\WlX.
The bulk-boundary propagator is given by
\eqn\bulkbdrypropa{
\langle \phi(x_0,y_0)\ i\partial_y\bar{\phi}(x_k)\rangle = \cL_y(x_0-x_k,y_0)\ ,
}
where the Lorentzian $\cL_y(x,y)$ is defined to be
\eqn\Lx{ \cL_y(x,y) \equiv \frac{y}{x^2 + y^2}\ .}
There are $j!$ contractions that are possible between the bulk operator
and the $j$ boundary insertions. As each integral gives a factor of $\pi$,
we are led to the following result after summing over all terms in the bulk
Landau-Ginzburg superpotential
\eqn\Aresultm{
\frac{\partial^2 \cW}{\partial \lambda \partial X} 
  = \sum_{j=0}^{k+2} g_{k+2-j}(t)\ \left(\frac{\pi X}{\sqrt2}\right)^{j} \ .
}
Finally, the last expression can be integrated to 
\eqn\Aresultp{
\cW =  \frac{\sqrt2\lambda}\pi\sum_{j=0}^{k+2} g_{k+2-j}(t)\ 
\frac{(\pi X/\sqrt2)^{j+1}}{j+1} \ .
}

The above computation may seem to be valid only when the coupling, $X$,
is a scalar and {\it not} a coupling matrix, $\cmat{X}$. Since then one
should treat the path-ordering carefully by also taking into account the
off-diagonal entries of $\cmat{X}$, which correspond to boundary
condition changing operators. However, a careful treatment of the
path-ordering, which comes into play when the boundary coupling, $X$, is
enhanced to a matrix, $\cmat{X}$, shows that the sole effect is taken
care of by the replacement
\eqn\XtoMatX{
X^{j+1}\longrightarrow {\rm Tr}(\cmat{X}^{j+1})\ . }
Then eq.~\Aresultp\ becomes in terms of the coupling matrix, $\cmat{X}$,
\eqn\Aresult{
\cW = \frac{\sqrt2\lambda}\pi\sum_{j=0}^{k+2} g_{k+2-j}(t)\ 
\frac{{\rm Tr}[(\pi\cmat{X}/\sqrt2)^{j+1}]}{j+1} \ .
}

\subsec{A second alternative computation}
\subseclab\secondcomp
We will now evaluate $\frac{\partial^3\cW}{\partial X^3}$ and verify
that the result is compatible with eq.~\Aresult.  The motivation for
carrying out this computation is two-fold: First, as we will see, the
analysis is not quite the same as in section~\firstcomp. Therefore, it
serves as a non-trivial check on the previous result for the effective
superpotential, $\cW$. Second, it provides for a simple example, which
allows us to illustrate the combinatorics involved in relating
correlators to effective superpotential terms. 

We identify $\frac{\partial^3\cW}{\partial X^3}$ with a three-point
function of three boundary operators, which, in order to fix the
$PSL(2,\IR)$ symmetry, are zero forms located at $x=0,1$ and $+\infty$.
Thus, we expect
\eqn\threepta{
\frac{\partial^3\cW}{\partial X^3} ={\bf 2}\  \Big\langle \bar{\xi}(0) \bar{\xi}(1) 
\bar{\xi}(+\infty)\Big\rangle\ .  }
The need for the factor of two in the above expression is subtle. The
simplest way to understand this is to study the precise relationship
between the first computation in section~\firstcomp, where one bulk
operator and one boundary operator were chosen as zero forms, and the
current computation, where three boundary operators are chosen as zero
forms. These two different choices can be related to each other by Ward
identities as was shown in refs.~\refs{\HofmanCE,\HerbstJP}.  From this
analysis it follows that we need to sum over two configurations, which
are obtained by exchanging the operators fixed at $0$ and $1$.\foot{%
The operator at $+\infty$ is identical in both situations.} Since the
operators at $0$ and $1$ are identical, there appears a factor of two in
the symmetrized correlator~\threepta.

As before the only non-vanishing contribution occurs for a
single bulk insertion, which is now a two-form operator and takes the form
(\cf Appendix A):
\eqn\Btwofrom{
V_j^{(2)}= \frac{j(j-1)}{2}\, g_{k+2-j}(t)\ \phi^{j-2}\, \tau\, \xi \ .
}
Further, the total $R$-charge constraint implies that we 
need to have $(j-2)$ integrated
boundary insertions.  The relevant fermionic free-field propagators are
\eqn\bulkbdrypropb{
\langle \tau(x,y) \bar{\xi}(w) \rangle = \cL_y(x-w,y)\ , \quad\quad
\langle \xi(x,y) \bar{\xi}(w) \rangle = \cL_x(x-w,y) \ ,
}
in terms of the Lorentzians
\eqn\Lxy{
\cL_y(x-w,y)\equiv \frac{y}{(x-w)^2 + y^2} \ , \quad\quad
\cL_x(x-w,y)\equiv \frac{x-w}{(x-w)^2 + y^2}\ . }
Then carrying out the various contractions, one obtains:
\eqn\threeptb{\eqalign{
\frac{\partial^3\cW}{\partial X^3} =&{\bf 2}\sum_j \frac{j(j-1)}{2}\ 
g_{k+2-j}(t)  \left(\frac{X}{\sqrt2}\right)^{j-2}
\int\limits_0^{+\infty} dy \int\limits_{-\infty}^{+\infty} dx
\left(\prod_{i=1}^{j-2}\int\limits_{-\infty}^{+\infty}\!\! dx_i \cL_x(x-x_i,y)\right) 
\cr
&\times \left(\cL_y(x-1,y)\,\cL_x(x,y) - \cL_x(x-1,y)\,\cL_y(x,y)\right)\ .
}}
The two terms in the second line arise from the two possible fermionic
contractions. The combinatoric factor $(j-2)!$ originates from the
number of bosonic contractions, which cancels the factor $(j-2)!$ in the
denominator. The latter factor appears from expanding the exponential in
the disk partition function~\diskpartfnb\ or in other words from
`pulling down' $(j-2)$ boundary insertions. Finally, the $(j-2)$
boundary integrations are easy to carry out and yield the factor
$\pi^{j-2}$, whereas the bulk integration contributes $\pi^2\over 2$. 
Putting all these numerical factors together, we arrive at
\eqn\Bresult{
\frac{\partial^3\cW}{\partial X^3} =\lambda \sum_j j(j-1) \frac{\pi^2}{2}g_{k+2-j}(t) 
\left(\frac{\pi X}{\sqrt2}\right)^{j-2} \ .
}
This result is clearly consistent with the effective
superpotential~\Aresult\ computed in the previous subsection.

\subsec{Proof of the HLL conjecture}

For $A$-minimal models at low values of $k$ Herbst, Lerche and
Lazaroiu~(HLL) explicitly solve the consistency conditions on
open-closed amplitudes such as the $A_\infty$-constraints, the
bulk-boundary crossing symmetry and the Cardy constraint
\refs{\HerbstJP,\HerbstZM}. Based on this analysis, they conjectured for
the $A$-minimal models a formula for the generating function of
tree-level open-string amplitudes. The HLL~formula precisely matches the
generating function~\Aresult\ computed in section~\firstcomp, and hence
this computation can be thought of as a proof of the HLL~conjecture. In
fact, the generating function~\Aresult\ is precisely the action for the
holomorphic matrix model considered in ref.~\HerbstZM.

Let us illustrate the connection to the HLL formula with a simple
example. For the $L=1$ boundary state of the $A$-minimal model at level
$k=3$ the explicit computation of HLL yields
\eqn\WHLL{\cW_{HLL}(u_1,u_2,t) = \sum_{j=0}^{5} g_{5-j}(t)\ h_{j+1}(u_1,u_2)\ , }
where the functions, $g_l(t)$ (with $g_1=0$), depending on the flat bulk
coordinates, $t$, are defined in eq.~\superone. The functions, $h_j(u)$,
are specific homogeneous functions of degree $j$, while the boundary
variables, $u_1$ and $u_2$, have assigned degrees~$1$ and $2$,
respectively. For example, the first few functions, $h_j$, are
\eqn\HLLhfunc{h_0=1 \ , \quad h_1=u_1 \ ,\quad h_2=u_2+\frac{u_1^2}2 \ , \quad
              h_3=u_1u_2 +\frac{u_1^3}3 \ , \quad \ldots \ . }
In order to compare the generating function~\Aresult\ to eq.~\WHLL\ we
evaluate the generating function~\Aresult\ at level $k=3$ and choose for
the coupling matrix, $\cmat{X}$, a $2\times 2$ matrix so as to model the
$L=1$ bound state. Then we find a precise agreement with the
superpotential $\cW_{HLL}$ if we set $h_j=\frac{\pi^j{\rm
Tr}(\cmat{X}^j)}{2^{j/2}j}$ and if we further identify the two
invariants of the coupling matrix, $\cmat{X}$, with the variables, $u_1$
and $u_2$:
\eqn\HLLident{u_1 \equiv {\pi\,{\rm Tr}(\cmat{X})\over\sqrt2} \ , \quad 
              u_2 \equiv -{\pi^2\,\det(\cmat{X})\over 2} \ . }
%

\newsec{The perturbative computation for the cubic torus}\seclab\perttorus
In this section we evaluate the topological disk partition function for
the `long' branes on the cubic torus. As before we carry out the
computation by using the perturbative techniques applied in the previous
section. However, compared to the $A$-type minimal model the analysis on
the torus is more involved since the whole disk partition function turns
out to be a series of an infinite number of different graphs. This
feature is due to the fact that the `long' branes, $L_a$, possess an
open-string modulus generated by a marginal open-string operator. We
evaluate only the first view terms in this series explicitly, but in
addition we are able to extract certain properties of the series to all
orders.

\subsec{The effective superpotential of the cubic torus} \subseclab\effWtorus
In order to obtain the effective superpotential, $\cW$, of the torus we
first compute the correlation function, $\frac{\partial^2\cW}{\partial
\lambda \partial \nu}$, which reads
\eqn\twopointb{
\frac{\partial^2\cW}{\partial \lambda \partial \nu}
=
\left\langle V_W^{(0)}(x_0,y_0)\,3!\,U\,{\bf \bar{\bxi}_1\bar{\bxi}_2\bar{\bxi}_3}(+\infty) \right\rangle \ . }
Recall that the parameters, $\lambda$ and $\nu$, are the couplings
arising in the bulk and boundary interactions~\genLGaction\ and \bintb,
and differentiating with respect to these parameters pulls down one bulk
and one boundary insertions. They are taken as the bulk zero form,
$V_W^{(0)}=W(\phi)$, at $(x_0,y_0)$ and the boundary zero form at
$+\infty$ in order to fix the $PSL(2,\IR)$ symmetry of the upper half
plane. This particular correlation function turns out to be a convenient
choice to saturate the three fermionic zero modes, $\bar{\xi}_i$,
$i=1,2,3$, which we have indicated in the correlators in bold face
letters.

Next, in the perturbative computation, we need to expand the
exponential~\diskpartfnb\ containing the bulk and boundary interactions
$S_W$ and $S_\partial$, where we keep only terms that satisfy the total
$R$-charge condition.  A simple analysis shows that the only
non-vanishing terms occur for three integrated $\psi^{(1)}$ insertions
and an equal but arbitrary number of integrated $V^{(2)}_W$ and
$\Omega^{(1)}$ insertions. Thus, unlike in the $A$-minimal model, we
find that the effective superpotential, $\cW$, as defined in
eq.~\diskpartfnb\ receives contributions from all terms, $\cW_n$. This
is due to the fact that both the boundary operator, $\Omega$, and the
bulk operator, $W(\phi)$, are marginal, which means that their
integrated one-form and two-form versions do not change the $R$-charge
of the correlator.

To summarize, the contribution to $\cW_n$ arises from $n$ bulk
insertion, $n$ $\Omega$~insertions and three $\Psi$~insertions. So the
perturbative computation, written in terms of free-field correlators,
leads to the following infinite sum after taking into account the total
$R$-charge condition:
\eqn\pertexpb{\eqalign{
\frac{\partial^2\cW}{\partial \lambda \partial \nu}
=& \sum_{n=1}^\infty \frac{(\lambda\nu)^{n-1}\mu^3}{[(n-1)!]^2}\times \cr
&\left\llangle V_W^{(0)} \left(\int_\Sigma\! V_W^{(2)}\right)^{n-1} \ 
P\left[ \left(\int_{\partial\Sigma}\! \psi^{(1)}\right)^3 \left(\int_{\partial\Sigma}\! \Omega^{(1)}\right)^{n-1}\right]
\cmat{U}\ {\bf \bar{\bxi}_1\bar{\bxi}_2\bar{\bxi}_3}(+\infty) \right\rrangle \ .
}}
In the above expression, all boundary integrals run over the full
$x$-axis as follows from the Dyson formula for a path-ordered
exponential.  As we demonstrate in Appendix~C the final result depends
only on the combination $(u_1+u_2+u_3)$, which is the trace of the
Chan-Paton matrix, $\cmat{U}$. So for computational simplicity, we can
set $u_1=u_2=u_3={u\over 3}$ such that the matrix, $\cmat{U}$, becomes
proportional to the identity matrix, ${\bf 1}_{3\times 3}$. Therefore
the path ordering, involving the boundary operators, $\Omega$, becomes
easier to handle. Putting in the explicit form of the various operators,
we obtain
\eqn\pertexpb{\eqalign{
\frac{\partial^2\cW}{\partial \lambda \partial \nu}
&=\frac{u}{3}\sum_{m=0}^\infty \frac{(3!\lambda\nu u)^{m}\mu^3}{[m!]^2}
\left(\frac{1}{\sqrt2}\right)^{3m+3}\times \cr
&\left\llangle [c\phi^3] \left(\int_\Sigma\!\! [c\phi\tau\xi]\right)^{n-1} \! 
P\left[ \left(\int_{\partial\Sigma}\!\! [\cmat{X}\partial_y\bar{\phi}]\right)^3\! 
\left(\int_{\partial\Sigma}\!\! [\epsilon\bar{\xi}\bar{\xi}\partial_y\bar{\phi}] 
\right)^{n-1}\right]
 {\bf \bar{\bxi}_1\bar{\bxi}_2\bar{\bxi}_3}(+\infty) \right\rrangle \ , 
}}
where we have used the short-hand notation $[c\phi^3]$ for
$c_{ijk}\phi^i\phi^j\phi^k$ and so on.
The next step in the computation is to carry out the various bulk-boundary
contractions and to evaluate the integrals 
\eqn\pertexpc{
\frac{\partial^2\cW}{\partial \lambda \partial \nu}
= \frac{u}{3}\sum_{m=0}^\infty \frac{(3!\lambda\nu u)^{m}\mu^3}{[m!]^2}
\left(\frac{\pi}{\sqrt2}\right)^{3m+3}\left(\frac{1}2\right)^m \sum_r
\cfa_{m+1,r}\ \gf_{m+1,r} \ ,
}
where every boundary integral gives a factor of $\pi$ and the bulk
integral gives a factor of $\pi^2\over2$ (\cf Appendix~C). The sum,
$\sum_r$, runs over all distinct contractions, which can be given a
graphical representation. The numerical coefficient, $\cfa_{n,r}$,
captures the the combinatorial multiplicity of the graph $r$ at order
$n$ in the parameter $u$. That is to say the coefficients, $\cfa_{n,r}$,
count the number of ways a given graph can be obtained. Finally, the
factor, $\gf_{n,r}$, describes the group-theoretic contribution of each
graph, which arises from summing over the couplings, $c_{ijk}$ and
$\epsilon^{ijk}$, of the marginal (bulk and boundary) vertex operators
and from tracing over the couplings, $\cmat{X}^i$, of the relevant
boundary operators.  The technical details concerning these issues are
collected in Appendix~B. 

Finally, we integrate eq.~\pertexpc\ to obtain the expansion in the effective
superpotential, $\cW$, 
\eqn\Bans{
\cW = \cW_0 + \frac1{3\gamma} \sum_{n=1}^{\infty} \frac{ \gamma^n u^n}{(n!)^2} 
      \sum_r \cfa_{n,r}\ \gf_{n,r} \ , }
where $\gamma\equiv 3(\pi/\sqrt2)^3$ and where we have set $\lambda=\mu=\nu=1$.
Obviously, the analyzed correlation function~\twopointb\ does not capture the
term, $\cW_0$, as it appears in eq.~\Bans\ as an integration constant. Therefore
we need to compute $\cW_0$ separately, which we turn to in the next subsection.

\subsec{Computing the zero-order term, $\cW_0$}

The correlation function~\twopointb\ considered in the previous subsection is
insensitive to the term $\cW_0$ of the expansion~\diskpartfnb.
Hence, in order to compute $\cW_0$, we consider now the correlation function,
$\frac{\partial^3 \cW}{\partial \mu^3}$, which is equal to the three-point
function of three boundary operators, $\psi^{(0)}$. As discussed in the context
of the $A$-type minimal model in section~\secondcomp, the precise identification
involves a factor of two:
\eqn\threeptb{
\frac{\partial^3\cW}{\partial \mu^3} ={\bf 2}\  \Big\langle \psi^{(0)}(0) \psi^{(0)}(1) 
\psi^{(0)}(+\infty)\Big\rangle\ . }
Since, we are only interested in the term, $\cW_0$, \ie the term with no
bulk and no $\Omega$~insertions, we evaluate the correlation
function~\threeptb\ at $u=0$ and obtain:
\eqn\threeptzero{\eqalign{
\left.\frac{\partial^3\cW_0}{\partial \mu^3}\right|_{u=0} 
&={\bf 2}\  \left\llangle 
\psi^{(0)}(0) \psi^{(0)}(1) 
\psi^{(0)}(+\infty)\right\rrangle \cr
&=2\ {\rm Tr}(\cmat X^i \cmat X^j \cmat X^k)\epsilon_{ijk} \left\llangle \bar{\xi}_1(0) \bar{\xi}_2(1)
\bar{\xi}_3(+\infty) \right\rrangle \cr
&=2\ {\rm Tr}(\cmat X^i \cmat X^j \cmat X^k)\epsilon_{ijk} \ . 
}}
Here the three $\bar{\xi}$ zero-modes are provided by the operator
insertions at $0,1$ and $+\infty$. Integrating the last expression and
setting $\mu=1$ we arrive at:
\eqn\Wzero{ 
\cW_0 = \frac13{\rm Tr}(\cmat X^i \cmat X^j \cmat X^k)\epsilon_{ijk}\ .  }
In fact, one can also compute $\cW_1$ starting from the correlation
function~\threeptb\ and verify that it agrees with the result obtained
in eq.~\Bans. However, we are not able to carry out the integrals that
appear in computing $\cW_n$ for $n>2$ from the correlation
function~\threeptb. Hence it is not possible to compare the higher order
terms.

\subsec{Gross features of the effective $D$-brane superpotential, $\cW$}\subseclab\grossfeatures
Several features of the structure of the effective $D$-brane
superpotential, $\cW$, may already be extracted without getting into
computational details. First of all, as previously discussed one can
obtain non-vanishing correlators only for the overall correct $R$-charge
and only for an equal number of bulk and $\Omega$~insertions. This
observation establishes the structure of the terms, $\cW_n$, in the
expansion~\diskpartfnb\ of $\cW$:
\eqn\struccW{
\cW_n \propto (g_0 u)^n {\rm Tr}(\cmat X^i \cmat X^j \cmat X^k)\ f_{ijk}\ . }
Here $f_{ijk}$ denotes a $SU(3)$ tensor. It is constructed from the
$n^{\rm th}$ symmetric power of the fully symmetric third-rank $SU(3)$
tensor, $c_{ijk}$, which enters in eq.~\struccW\ through the $n$ bulk
insertions. Further, the cyclic property of the trace implies that ${\rm
Tr}(\cmat X^i \cmat X^j \cmat X^k)$ must either be proportional to
$\epsilon^{ijk}$, which transforms as a $SU(3)$ singlet, or must be
again a symmetric third-rank tensor of $SU(3)$.  Thus this simple
group-theoretic analysis tells us that $\cW_n$ can only be non-zero if
the symmetric tensor product, $S^n(\tbox)$, contains either a singlet or
a symmetric third-rank tensor, $\tbox\,$.\foot{We abbreviate the
symmetric third-rank tensor of $SU(3)$ by its Young tableau, $\tbox\,$.}

By decomposing the trace, ${\rm Tr}(\cmat X^i \cmat X^j \cmat X^k)$,
into its $SU(3)$ representations and by comparing with the
representations appearing in $S^n(\tbox)$ we can even determine which
parts of the trace can possibly appear in the term, $\cW_n$. Expanding
${\rm Tr}(\cmat X^i\cmat X^j\cmat X^k)$, it is useful to reorganize it
into three kinds of terms, which we will call $\kappa_{111}$,
$\kappa_{123}$ and $\kappa_{132}$, and which are defined as
\eqn\Defkappa{\eqalign{
  \kappa_{111}(\cmat X) &={1\over 3}\sum_i{\rm Tr}(\cmat X^i\cmat X^i\cmat X^i)
     = x^1_{12}x^1_{23}x^1_{31}+x^2_{12}x^2_{23}x^2_{31}+x^3_{12}x^3_{23}x^3_{31}\ ,\cr
  \kappa_{123}(\cmat X) &={\rm Tr}(\cmat X^1\cmat X^2\cmat X^3)
     = x^1_{12}x^2_{23}x^3_{31}+x^3_{12}x^1_{23}x^2_{31}+x^2_{12}x^1_{23}x^3_{31}\ ,\cr
  \kappa_{132}(\cmat X) &={\rm Tr}(\cmat X^1\cmat X^3\cmat X^2)
     = x^1_{12}x^3_{23}x^2_{31}+x^2_{12}x^1_{23}x^3_{31}+x^3_{12}x^1_{23}x^2_{31} \ . }}
Then the combination $(\kappa_{123}-\kappa_{132})$ forms the $SU(3)$
singlet, while $\kappa_{111}$ and $(\kappa_{123}+\kappa_{132})$ are
components of the $SU(3)$ representation $\tbox\,$. Due to the special
structure of the bulk couplings, $c_{ijk}$, given in eq.~\Defc, only
these two components appear in the effective superpotential, $\cW$.

First, we observe that the reducible representation, $S^2(\tbox)$, contains
neither a singlet nor the representation $\tbox\,$, and therefore
we conclude $\cW_2=0$. Further, using the computer algebra package LiE \LiE,
we have checked up to $n=20$, that singlets appear only in the decomposition of
$S^{2n}(\tbox)$ while the representation $\tbox\,$ arise only
in $S^{2n+1}(\tbox)$. In the next section we present 
an argument that shows that this pattern is indeed true to all orders $n$.
Note also that the first instance when the multiplicities in $S^k(\tbox)$ 
of the two representations in question is greater than one occurs for the
first time at $k=7$.

Finally, up to an overall proportionality constant, we can write the first
six terms in $\cW$ by explicitly working out the $f_{ijk}$, and we find:
\eqn\Wn{\eqalign{
  \cW_0 & = \cI_0\left(-3\left(\kappa_{123}-\kappa_{132}\right)\right) \ , \cr
  \cW_1 & = \cI_1\left(3\,\kappa_{111}
          -{3\over 2}a\left(\kappa_{123}+\kappa_{132}\right)\right)g_0u \ , \cr
  \cW_2 & = 0 \ , \cr
  \cW_3 & = \cI_3\left(-{9\over 2}a^2 \kappa_{111}+\left(3-{3\over 4}a^3\right)
          \left(\kappa_{123}+\kappa_{132}\right)\right)\left(g_0u\right)^3 \ , \cr
  \cW_4 & = \cI_4\left(-{9\over 2}a^4-36 a\right)\left(\kappa_{123}-\kappa_{132}\right)
          \left(g_0u\right)^4 \ , \cr
  \cW_5 & = \cI_5\left(\left({3\over 8}a^4+3 a\right)\kappa_{111}
          -\left({3\over 16}a^5+{3\over 2}a^2\right)\left(\kappa_{123}+\kappa_{132}\right)\right)
          \left(g_0u\right)^5 \ , \cr
  \cW_6 & = \cI_6\left({9\over 4}a^6-45a^3-18\right)\left(\kappa_{123}-\kappa_{132}\right)
          \left(g_0u\right)^6 \ , }}
with 
\eqn\CoeffcI{\cI_0={1\over 3} \ , \quad \cI_1 = {2\gamma\over 3} \ ,\quad 
             \cI_3=-{4\gamma^3\over 9} \ , \quad \cI_4=-{2 \gamma^4\over 9} \ , }
where $\gamma\equiv 3(\pi/\sqrt2)^3$. In Appendix~B, we discuss the
details of this computation. But we want to emphasis here that the
group-theoretical structure, \ie the appearance of the appropriate
traces~\Defkappa\ and the vanishing of the term, $\cW_2$, arises also
directly by explicitly evaluating the correlation function~\twopointb\
as outlined in section~\effWtorus\ and in Appendix~B.

\newsec{Modular properties of the toroidal effective superpotential}

In section~\perttorus\ we perturbatively computed the effective
superpotential, $\cW$, for the `long' branes, $L_a$, of the
Landau-Ginzburg model with the cubic superpotential~\Wtorus. In this
analysis, however, we have not really used the geometry of the
underlying torus, $\cT$. Thus, in this section we exploit the modular
properties of the torus, $\cT$, in order to extract further properties
of the effective superpotential, $\cW$.

It is well-known that the Landau-Ginzburg theory~\Wtorus\ corresponds in
the large radius limit of the K\"ahler moduli space to a supersymmetric
$\sigma$-model, for which the target space is the torus, $\cT$, given as
the hypersurface, $W = 0$, in the projective space, $\IP^2$ \WittenYC.
The parameter, $a$, in the superpotential, $W$, parametrizes the complex
structure of this hypersurface, and it is related to the standard
complex structure modulus, $\tau$, of the torus via the relation~\Defa.
Note that for a torus with complex structure, $\tau$, there are
generically twelve different possible values for the parameter, $a$,
which are the roots of the order twelve polynomial associated to
eq.~\Defa. All these distinct roots describe in the large radius limit
identical $\sigma$-models, and hence the associated Landau-Ginzburg
theories are also equivalent. 

In our setup, where we treat the Landau-Ginzburg superpotential~\Wtorus\
perturbatively, different ratios~\Moda\ of the couplings $g_0$ and
$g_1$, yet ratios associated via eq.~\Defa\ to the same modulus, $\tau$,
should also give rise to equivalent correlation functions.\foot{At least
as long as the couplings~$g_0$ and $g_1$ are small.} In particular, as
the boundary operator, $\Omega$, is independent of the structure of the
perturbative superpotential~\Wtorus, the superpotential contributions,
$\cW_n$, which are graded by the number of $\Omega$ insertions, must
also be related order by order for equivalent Landau-Ginzburg theories.

In order to study the properties of the superpotential terms, $\cW_n$,
we rephrase the question in an appropriate language. First of all, the
relationship~\Defa\ defines the parameter, $a$, to be a modular function
of $\Gamma[3]$, \ie the function, $a$, is invariant under the action of
the group~$\Gamma[3]$. Moreover, the different roots of the polynomial
associated to eq.~\Defa\ are permuted under the Galois group of the
polynomial. The Galois group of this polynomial is the tetrahedral group
$\cT_{12}\simeq PSL(2,\IZ_3)\simeq PSL(2,\IZ)/\Gamma[3]$ of index~$12$.
From this perspective using the properties of the modular function, $a$,
and by requiring invariance of the correlators, $\cW_n$, we want to
determine the modular transformation behavior of the Chan-Paton traces,
$\kappa_{111}$, $(\kappa_{123}+\kappa_{132})$,
$(\kappa_{123}-\kappa_{132})$, and of the coupling product, $g_0u$.

We study the group, $\cT_{12}$, by looking at its generators, $S$ and
$T$, which act upon the complex structure, $\tau$, as\foot{Since the
group $PSL(2,\IZ)$ is generated by the group elements $S$ and $T$, also
the tetrahedral group, $\cT_{12}$, as a quotient group of $PSL(2,\IZ)$,
is generated by $S$ and $T$.} 
\eqn\STgenerators{S:\ \tau \to -{1\over\tau} \ , \quad\quad T:\ \tau\to\tau+1 \ .}
Then eq.~\Defa\ encodes the transformation behavior of the modular
function, $a$, to be \LercheCS
\eqn\STa{S:\ a \to {a+2\over a-1} \ , \quad\quad T:\ a\to \rho^2\,a \ , }
where $\rho=e^{2\pi i\over 3}$. It is straight forward to check that for
a given root, $a$, these two transformations generate all the other
roots of the polynomial associated to eq.~\Defa.

The next task is to deduce the modular properties of the remaining
quantities. From the superpotential term, $\cW_0$, in eq.~\Wn\ we
readily see that the Chan-Paton trace, $(\kappa_{123}-\kappa_{132})$, is
invariant under the group, $\cT_{12}$, \ie
\eqn\STksing{S:\ \kappa_{123}-\kappa_{132} \to \kappa_{123}-\kappa_{132} \ , \quad\quad
             T:\ \kappa_{123}-\kappa_{132} \to \kappa_{123}-\kappa_{132} \ . }
This allows us directly to deduce from the term, $\cW_4$, in eq.~\Wn\ the group
action on the product, $g_0 u$:
\eqn\STgu{S:\ g_0 u \to -i{(a-1)\over\sqrt3} \,g_0u \ ,
          \quad\quad T:\ g_0u \to \rho^{-2}\,g_0u \ . }
Finally, we determine from the superpotential contribution, $\cW_1$, the
modular properties of the Chan-Paton traces, $\kappa_{111}$ and
$(\kappa_{123}+\kappa_{132})$,
\eqn\STkappa{\eqalign{
   S:&\ \pmatrix{\kappa_{111}\cr\kappa_{123}+\kappa_{132}} \to
       \pmatrix{{i\over\sqrt3}\left(\kappa_{111}+\left(\kappa_{123}+\kappa_{132}\right)\right)\cr
       {i\over\sqrt3}\left(2\kappa_{111}-\left(\kappa_{123}+\kappa_{132}\right)\right)} \ , \cr
   T:&\ \pmatrix{\kappa_{111}\cr\kappa_{123}+\kappa_{132}} \to
       \pmatrix{\rho^2\,\kappa_{111} \cr \kappa_{123}+\kappa_{132} } \ . }}
Note that in this analysis we have only used the terms $\cW_0$, $\cW_1$
and $\cW_4$ in eq.~\Wn\ in order to arrive at the transformation
rules~\STksing, \STgu\ and \STkappa. The other terms in the list \Wn\
serve as non-trivial checks and confirm the stated results. 

On the other hand we can also use the derived modular properties so as
to constrain the general structure of the superpotential terms, $\cW_n$.
In particular we now show that the Chan-Paton trace,
$(\kappa_{123}-\kappa_{132})$, does only appear in $\cW_n$ for even values
of $n$. As discussed in the previous section a contribution to $\cW_n$
involving $(\kappa_{123}-\kappa_{132})$ has the general structure
\eqn\Weven{\left(\sum_{k=0}^n \alpha_k (a-1)^k\right)\,\left(g_0 u\right)^n
                 \,\left(\kappa_{123}-\kappa_{132}\right) \ , }
with numerical coefficients, $\alpha_k$. Invariance of this expression with 
respect to the generator, $S$, constrains the coefficients, $\alpha_k$:
\eqn\Wevencon{0=\sum_{k=0}^n {\alpha_k\over(\sqrt3)^{n-k}}
              \left((\sqrt3)^{n-k}(a-1)^k+i^n(\sqrt3)^k(a-1)^{n-k}\right) \ . }
Note that in this formula $i^n$ becomes $\pm 1$ for even integers, $n$,
which implies that the polynomials of the coefficients, $\alpha_k$ and
$\alpha_{n-k}$, in eq.~\Wevencon\ are linearly dependent. Therefore the
condition~\Wevencon\ can always be non-trivially fulfilled for even
values of $n$. However, for odd integers, $n$, we get $i^n=\pm i$, and
as a consequence the polynomials of all the coefficients, $\alpha_k$, in
eq.~\Wevencon\ are linearly independent, which enforces all the
coefficients, $\alpha_k$, to be zero. Thus the Chan-Paton trace,
$(\kappa_{123}-\kappa_{132})$, can never contribute to odd orders in the
effective superpotential, $\cW$. By similar arguments one also shows
that the other Chan-Paton traces, $\kappa_{111}$ and
$(\kappa_{123}+\kappa_{132})$, do not appear at even orders in the
effective superpotential, $\cW$.\foot{If we take into account the
transformation behavior with respect to the generator, $T$, we find that
the term, $\cW_2$, must also vanish by modularity.} Hence this analysis
confirms to all orders in $n$ the group-theoretical claims made in
section~\grossfeatures\ and we conclude that the effective
superpotential, $\cW$, possess the $\IZ_2$ symmetry
\eqn\Wsym{\big(u,\kappa_{111}(\cmat X),\kappa_{123}(\cmat X),
\kappa_{132}(\cmat X)\big)\to
  \big(-u,-\kappa_{111}(\cmat X),-\kappa_{132}(\cmat X),-\kappa_{123}(\cmat X)\big) \ . }

In order to gain further insight into the meaning of the open-string
couplings, $u$, we want to reinterpret the modular behavior of the
product, $g_0u$. As we have discussed in section~\bulktorus\
to describe the deformations of Landau-Ginzburg superpotential,
$W$, in terms of flat coordinates, we need to identify the coupling
constant, $g_0$, with the inverse `flattening' factor, $\qf^{-1}$, 
which we introduced in eq.~\Defqf.
From the transformation behavior of the modular functions, $a$, in
eq.~\STa, we immediately deduce for the `flattening' factor~\Defqf\
(and, hence, also for the inverse coupling, $g_0^{-1}$)
\eqn\STqf{
  S:\ \qf\to {\sqrt3i\over\tau(a-1)} \qf \ , \quad\quad
  T:\ \qf\to \rho^2 \qf \ . }
Comparing with eq.~\STgu\ we identify the coupling, $g_0$, with the
inverse factor, $\qf^{-1}$, and then we deduce for the open-string
parameter, $u$, the modularity:
\eqn\STu{S:\ u\to {u\over\tau} \ , \quad\quad T:\ u\to u \ . }
Note that these transformations of the open-string coupling, $u$, match
the modular properties of a point on the torus, $\cT$. 

Let us pause to stress the significance of this result. The perturbative
treatment of the Landau-Ginzburg superpotential contains naturally the
`flattening' factor, $\qf$, which in ref.~\LercheWM\ is determined by
differential equations arising from the periods of the torus. Moreover,
due to eq.~\STu\ the open-string coupling, $u$, can be thought of as a
point on the torus. Thus this parameter is the (combined) open-string
modulus of the three `long' branes, $L_a$, and it coincides with the
flat open-string coordinate, $u$, used in
refs.~\refs{\BrunnerMT,\GovindarajanIM}.\foot{In principle by the
presented arguments the coupling, $u$, could still differ from the flat
open-string coordinate by a multiplicative factor given by a modular
invariant function. However, in the sequel we will show that the
coupling, $u$, is indeed the flat coordinate.} 

Furthermore, as the parameter, $u$, transforms as point on the torus and
as the Chan-Paton trace, $(\kappa_{123}-\kappa_{132})$, is according to
eq.~\STksing\ modular invariant, the $\tau$-dependent part in the even
superpotential terms, $\cW_{2k}$, transform as modular functions,
$\tilde G_{2k}$, of weight $2k$,\foot{Due to their modular properties
the functions, $\tilde G_{2k}$, are proportional to the Eisenstein
functions, $G_{2k}$, for $2\le k\le 5$.  However, this need {\it not} be
true for $k>5$. Since the Eisenstein functions, $G_4$ and $G_6$,
generate all the higher weight Eisenstein functions, for instance only a
particular linear combination of $G_4^3$ and $G_6^2$ coincides with the
Eisenstein function, $G_{12}$, which might not be proportional to
$\tilde G_{12}$. Finally, it is interesting to note that the
multiplicity of the singlets in the symmetric tensor product,
$S^{2k}(\tbox)$, as discussed in section~\grossfeatures, coincides (at
least for $2\le k\le 10$) with the number of linearly independent
modular functions of weight~$2k$.} \ie
\eqn\evenW{\cW_{2k} = \tilde G_{2k}(\tau)\,\left(\kappa_{123}-\kappa_{132}\right) u^{2k} \ , }
with
\eqn\ModFunc{\tilde G_{2k}\left(a\tau + b \over c\tau+ d\right) =
                    \left(c\tau+d\right)^{2k} \tilde G_{2k}(\tau) \ . }

Before concluding this section we rewrite the effective superpotential,
$\cW$, so as to explicitly show its general structure and its dependence
on the bulk and boundary moduli, $\tau$ and $u$:
\eqn\rewriteB{\cW(\tau,u,\cmat X) = 
    \Delta_{111}(\tau,u)\,\kappa_{111}(\cmat X) +\Delta_{123}(\tau,u)\,\kappa_{123}(\cmat X) 
    +\Delta_{132}(\tau,u)\,\kappa_{132}(\cmat X) \ . }
Note that due to the symmetry property \Wsym\ of the effective
superpotential, $\cW$, the functions, $\Delta_{ijk}$, obey
\eqn\symDB{\Delta_{111}(\tau,u) = - \Delta_{111}(\tau,-u) \ , \quad\quad
           \Delta_{123}(\tau,u) = - \Delta_{132}(\tau,-u) \ . }
%

\newsec{Mirror map and disk instantons for the `long' A-branes on the torus}
Up to now we have performed all our computations in the topological
B-model. In this section we compute the $D$-brane effective
superpotential of the configuration, which is mirror symmetric to the
`long' branes of the cubic torus, $\cT$. By comparing the effective
superpotential of the A-model with the one of the B-model we are able
deduce the open-string mirror map. Since the two effective
superpotentials are comprised of a collection of correlation functions,
the existence of a unique mirror map is also an indirect but non-trivial
check on the perturbative B-model computations performed in the previous
sections.

\subsec{The effective superpotential in the A-model}
All contributions to the effective $D$-brane superpotential, $\hat\cW$,\foot{%
We distinguish the A-model from B-model quantities
by adding a hat~$\hat{}\,$ for A-model quantities.}
on the A-model side
arise from non-perturbative effects, namely from worldsheet disk
instantons \KachruAN. In general summing up the contributions of those
disk instantons in a Calabi-Yau manifold is a highly non-trivial
problem. However, since the mirror manifold of the torus, $\cT$, is
yet again a torus, $\hat{\cT}$, the special Lagrangian submanifolds,
which represent $A$-type $D$-branes in the geometric regime of the A-model,
are just given by real lines on the torus~$\hat{\cT}$. Hence
analyzing disk instantons for this toroidal geometry simply amounts to
computing areas of triangles, which has been carried out in
refs.~\refs{\PolishchukDB,\CremadesQJ,\BrunnerMT,\HerbstNN}.\foot{For
superpotential terms involving more than three boundary changing
operators one needs to compute areas of polygons \HerbstNN.} In order to
set the stage for the next section we briefly review this analysis for
the mirror branes, $\hat L_a$, of the `long' branes, $L_a$. 

\figinsert\lbranes{The torus, $\hat\cT$, is shown together with the
three $A$-type D1-branes~$\hat L_a$. Their real moduli, $\hat\beta_a$,
parametrize the offsets of the branes, $\hat L_a$, in the horizontal
direction. The boundary changing operators, $\hat x_{ab}^i$, are located
at the intersection of the branes, $\hat L_a$ and $\hat
L_b$.}{2.3in}{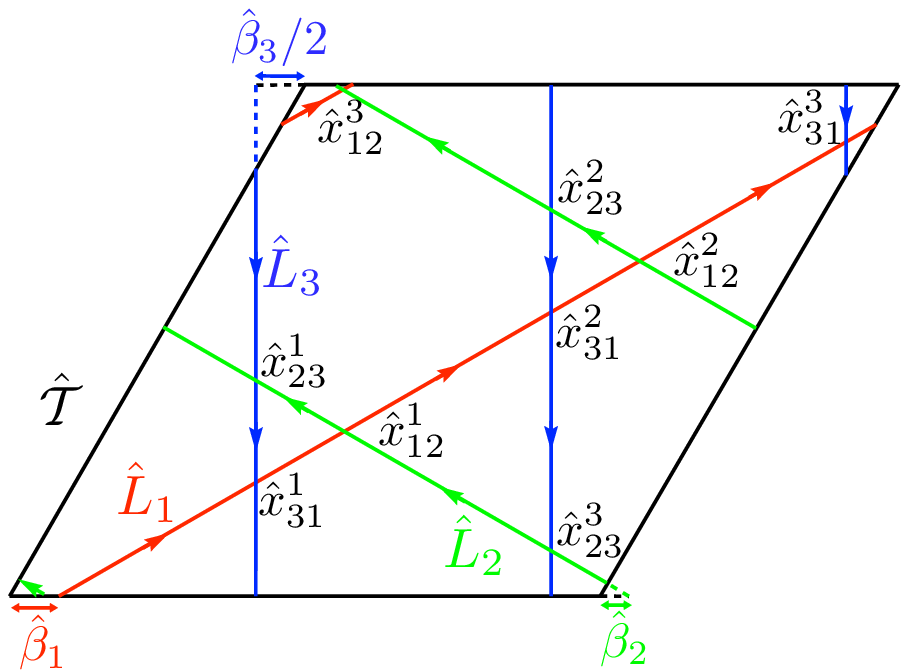}

On the B-model side we have performed our computations at the Landau-Ginzburg
point, which is located at the point~$\rho = e^{{2\over 3}i\pi}$ in the
K\"ahler moduli space. As mirror symmetry exchanges the K\"ahler and the
complex structure moduli, $(\tau,\rho)\leftrightarrow(\hat\rho,\hat\tau)$,
of mirror pairs, we consider the mirror torus, $\hat\cT$, at $\hat\tau =
e^{{2\over 3}i\pi}$ in the complex structure moduli space. Moreover, we
need to identify the A-model mirror images, $\hat L_a$, of the B-model
`long' branes, $L_a$. At the Landau-Ginzburg point the three `long'
branes, $L_a$, carry the RR-charges~\refs{\BrunnerMT,\GovindarajanIM}
\eqn\Lcharges{L_a:\quad (r,c_1)^{\rm LG} \ = \ \left\{ (-1,1),\,(-1,-2),\,(2,1) \right\} \ , }
where the charge vector~$(r,c_1)$ denotes the $D2$-brane and $D0$-brane
charge respectively. On the A-model side these charges become the
winding numbers of the corresponding one-dimensional Lagrangian
submanifolds of the mirror torus, $\hat\cT$ \PolishchukDB. This allows us
to identify the A-model branes, $\hat L_a$, depicted in \lfig\lbranes.
Furthermore the boundary changing operators, $\hat x^i_{ab}$, are located
at the intersections of the lines associated to the branes, $\hat L_a$
and $\hat L_b$ .\foot{The orientation of the intersection specifies
whether the corresponding boundary changing operator is fermionic or
bosonic. In the following we concentrate only on fermionic boundary
changing operators as their couplings appear in the effective
superpotential, $\hat{\cW}$.}

The A-model correlators are given by the sum of disk instantons bounded
by the associated Lagrangian submanifolds and weighted by their
(complexified) areas. These areas depend on the (real)
positions, $\hat\beta_a$, of the bounding branes, $\hat L_a$, and therefore
the three possible correlation functions take schematically the
form \PolishchukDB:
\eqn\Acorr{  
     C^A_{ijk}(\hat\rho,\hat\alpha_a,\hat\beta_a) 
       = \sum_k e^{2\pi i \hat\rho\, A_{ijk}^{(k)}(\hat\beta_a) }
                 e^{2\pi i \, W_{ijk}^{(k)}(\hat\alpha_a) } \ . }
Here the coefficients, $A^{(k)}_{ijk}$, denote the (dimensionless) areas
of the different disk instantons labeled by the index~$k$, whereas the
phases, $W^{(k)}_{ijk}$, are Wilson line contributions. The latter are
obtained by integrating the flat connection along the circumference of
the disk instanton. The connection of the brane, $\hat L_a$, is
parametrized by the single real modulus, $\hat\alpha_a$, because the
topology of the associated one-dimensional submanifold is a circle.
Finally, note that only the correlators $C^A_{111}$, $C^A_{123}$,
$C^A_{132}$ (and their cyclic permutations) are non-vanishing.

\figinsert\lbtriangle{For the $D1$-branes, $\hat L_1$, $\hat L_2$ and $\hat L_3$,
the non-trivial correlators~$\corr{\hat x^1_{31}\hat x^1_{12}\hat x^1_{23}}$, 
$\corr{\hat x^1_{31}\hat x^2_{12}\hat x^3_{23}}$ and 
$\corr{\hat x^1_{12}\hat x^3_{23}\hat x^2_{31}}$ arise from disk instantons. The figures 
show the disk instantons~$k=0$ for these three correlation functions
respectively.}{1.2in}{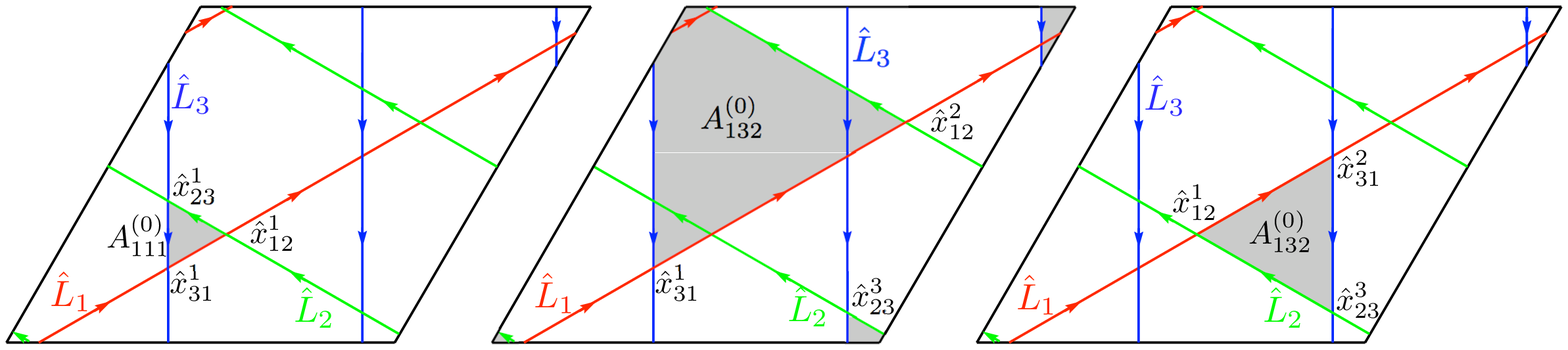}
The areas, $A^{(k)}_{ijk}$, of the disk instantons can be deduced by
evaluating the areas of all triangles formed by the corresponding
boundary changing operators in the correlators, $C^A_{ijk}$. The disk
instantons labeled by $k=0$ are shown for the non-vanishing correlators
in \lfig\lbtriangle.  A view steps of elementary geometry reveal for a
general instanton labeled by~$k\in\IZ$
\eqn\areas{
  A_{111}^{(k)} = {3\over 2}\left(k+{\hat\beta\over 3}\right)^2 \ , \quad
  A_{123}^{(k)} = {3\over 2}\left(k+{1\over 3}+{\hat\beta\over 3}\right)^2 \ ,\quad
  A_{132}^{(k)} = {3\over 2}\left(k-{1\over 3}+{\hat\beta\over 3}\right)^2 \ , }
and for its Wilson line contributions
\eqn\Wilson{
  W_{111}^{(k)} = \left(k+{\hat\beta\over 3}\right) \hat\alpha \ , \quad
  W_{123}^{(k)} = \left(k+{1\over 3}+{\hat\beta\over 3}\right) \hat\alpha \ , \quad
  W_{132}^{(k)} = \left(k-{1\over 3}+{\hat\beta\over 3}\right) \hat\alpha \ . }
Here we have used the
definitions~$\hat\alpha=\hat\alpha_1+\hat\alpha_2+\hat\alpha_3$ and
$\hat\beta=\hat\beta_1+\hat\beta_2+\hat\beta_3$. Inserting eqs.~\areas\
and \Wilson\ into the correlator~\Acorr\ we finally arrive in terms of
the complex variables $\hat\rho$ and $\hat u=\hat\alpha +
\hat\rho\,\hat\beta$ at
\eqn\AcorrExpl{C^A_{ijk}(\hat\rho,\bar{\hat\rho},u,\bar{\hat u}) = 
         \left(e^{{2\over 3}\pi i \hat\alpha\hat\beta}  \hat q^{\hat\beta^2\over 6} \right) \cdot
         \Delta^A_{ijk}(\hat\rho,u) \ ,}
with
\eqn\ADelta{\eqalign{
  \hat\Delta_{111}(\hat\rho,\hat u) &= \sum_{k\in\IZ} \hat q^{{3\over 2}
           k^2} e^{2\pi i k \hat u } \ , \cr
  \hat\Delta_{123}(\hat\rho,\hat u) &= \sum_{k\in\IZ} \hat q^{{3\over 2}
           \left(k+{1\over 3}\right)^2} e^{2\pi i \left(k+{1\over 3}\right) \hat u } \ , \cr
  \hat\Delta_{132}(\hat\rho,\hat u) &= \sum_{k\in\IZ} \hat q^{{3\over 2}
           \left(k-{1\over 3}\right)^2} e^{2\pi i \left(k-{1\over 3}\right) \hat u } \ , }}
and the non-holomorphic prefactor
\eqn\prefact{\pf = e^{{2\over 3}\pi i \hat\alpha\hat\beta}  \hat q^{\hat\beta^2\over 6} \ .}
The non-holomorphic prefactor, $\pf$, will be discussed in
subsection~7.3. Here, we only observe that this factor cannot appear
in the holomorphic superpotential and hence the A-model superpotential, $\hat\cW$,
is only given in terms of the holomorphic parts~\ADelta:
\eqn\Amodel{
   \hat\cW(\hat\rho,\hat u,\cmat{\hat X}) = 
       \hat\Delta_{111}(\hat\rho,\hat u) \kappa_{111}(\cmat{\hat X})
     +\hat\Delta_{123}(\hat \rho,\hat u) \kappa_{123}(\cmat{\hat X}) 
     +\hat\Delta_{132}(\hat\rho,\hat u) \kappa_{132}(\cmat{\hat X}) \ .  }
Here the matrices, $\cmat{\hat X}$, denote the A-model analogs of the B-model
coupling matrices, $\cmat X$, defined in eq.~\chanpaton.

\subsec{The mirror map and flat coordinates}
The aim of this subsection is to determine the mirror map for the `long'
branes, $L_a$, on the torus, $\cT$, and the mirror $D1$-branes, $\hat
L_a$, on the mirror torus, $\hat\cT$. That is to say we construct the
map between the sets of A-model and B-model variables, or, equivalently,
we determine the flat coordinates on the B-model side, which are
canonically identified with variables of the A-model. 

In the B-model the variables, which arise from the `long' branes, $L_a$,
on the torus, $\cT$, are the bulk complex structure modulus, $\tau$, the
collective $D$-brane modulus, $u$, and the coupling matrices, $\cmat X$, of the
boundary changing operators, $x_{ab}^i$.  As discussed in the previous
section, the natural variables for the mirror A-model are the bulk
K\"ahler modulus, $\hat\rho$, of the torus, $\hat\cT$, and the
complexified position modulus, $\hat u$, and the boundary changing
matrices, $\cmat{\hat X}$, of $D1$-branes, $\hat L_a$.

In the closed-string sector the complex structure modulus, $\tau$, is
the flat coordinate of the B-model bulk theory \LercheWM, and hence
mirror symmetry identifies the A-model K\"ahler modulus, $\hat\rho$,
with the B-model complex structure modulus, $\tau$. In the following for
ease of notation we will often replace the A-model K\"ahler parameter,
$\hat\rho$, by its mirror variable, $\tau$.

In the open-string sector we still must determine the mirror map, for
which we make the following ansatz,
\eqn\openansatz{
   \hat u = \cN_u(\tau)\ u + u_0(\tau) \ , \quad\quad
   \cmat{\hat X} = \cN_X(\tau,u) \ \cmat X \ . }
This ansatz is justified by two observations. First of all the
open-string mirror map should not contain any dimensionful couplings.
Therefore, the functional dependence between the variables, $\hat u$ and
$u$, can only involve (on the B-model side) the moduli, $u$ and $\tau$,
but not the dimensionful matrices, $\cmat X$. Furthermore, the additive
property of both the A-model variable, $\hat u$, and the B-model
variable, $u$, implies that $\hat u$ and $u$ are related linearly to one
another. Similarly, by dimensional arguments the coupling matrices,
$\cmat{\hat X}$ and $\cmat X$, can only be proportional to one another
with the moduli-dependent proportionality constant, $\cN_X(\tau,u)$.

Our first task is to determine the the shift $u_0(\tau)$. This is
achieved by adjusting the parameter~$u_0$ such that the symmetry
properties~\symDB\ of the functions~$\Delta_{ijk}(u)$ coincide with
those of $\hat\Delta_{ijk}(\hat u-u_0)$. That is to say for the functions,
$\mu_\ell$, defined by
\eqn\defmu{
   \mu_1(\tau,\hat u) = \hat\Delta_{111}(\tau,\hat u-u_0) \ , \quad
   \mu_2(\tau,\hat u) = \hat\Delta_{123}(\tau,\hat u-u_0) \ , \quad
   \mu_3(\tau,\hat u) = \hat\Delta_{132}(\tau,\hat u-u_0) \ , }
we require analogously to eq.~\symDB\ to obey the symmetries
\eqn\signperm{
    \mu_1(\tau,\hat u)=-\mu_1(\tau,-\hat u)\ , \quad\quad 
    \mu_2(\tau,\hat u)=-\mu_3(\tau,-\hat u) \ . }
It is easy to check that these two conditions are simultaneously
fulfilled for
\eqn\shift{u_0(\tau) = -{1\over 2} \tau - {1\over 2} \ . }
Note that the the shift, $u_0(\tau)$, required to match the A-model
variables with the B-model variables, has already been observed in
ref.~\BrunnerMT.

Equating the A-model superpotential, $\hat\cW$, with the B-model
superpotential, $\cW$, yields with with the ansatz~\openansatz\ for the
mirror map:
\eqn\mirrorrel{\cN_X^3(\tau,u)\ \hat\cW(\tau,\cN_u(\tau)u+u_0,\cmat X) 
   = \cW(\tau,u,\cmat X) \ . }
This relation allows us now to determine the open-string mirror map
functions, $\cN_u(\tau)$ and $\cN_X(\tau,u)$, explicitly. The technical
details of this computation are relegated to Appendix~D, where we derive
for the functions, $\cN_u(\tau)$ and $\cN_X(\tau,u)$, the expressions
\eqn\MirrorNu{\cN_u={3\,\cI_1\over\sqrt{2\pi i}\,\cI_0}={9 \sqrt{-i\pi^5}\over 2} \ , }
and 
\eqn\MirrorNX{\cN_X(\tau,u)=\root{3\,}\of{i\over\eta(\tau)}
              \ e^{-{\cN_u^2 \over 18}\,G_2(\tau)\,u^2+O(u^4)}   \ . }
The higher order terms in $u$ are determined by the precise numerical
coefficients of the superpotential terms, $\cW_{2k}$, in eq.~\evenW.
Here $G_2(\tau)$ is the second function in the Eisenstein series and it
is also given by
\eqn\Gtwo{G_2(\tau)= - 4 \pi i \,{\eta^\prime(\tau)\over\eta(\tau)} \ . }

Note that determining the open-string mirror maps, $\cN_u$ and $\cN_X$,
form eq.~\mirrorrel\ is a highly over-determined problem (\cf
Appendix~D). Hence the existence of the solution~\MirrorNu\ and
\MirrorNX\ is a non-trivial check on the method of computing the
effective superpotential, $\cW$, perturbatively.

\subsec{Holomorphic anomaly}

In the physical theory the non-holomorphic prefactor arises from the
K\"ahler potential and hence is a D-term contribution \CremadesQJ. Here
we want to interpret this prefactor from the topological A- and B-model
point of view. 

In the bulk theory of the B-model the correlators receive
non-holomorphic contributions due to the holomorphic anomaly, which
arises at higher genus amplitudes \BershadskyCX. Therefore in the
B-model with boundaries we similarly expect the appearance of
non-holomorphic terms. Since we have computed tree-level disk diagrams,
the obtained correlators are holomorphic in the moduli. However, due to
quantum $A_\infty$ relations the disk diagrams are ultimately linked to
open one-loop cylinder amplitudes \refs{\HerbstKT,\HerbstNN}, which
potentially suffer from a holomorphic anomaly. But to clarify the
precise r\^ole of non-holomorphic terms in the quantum $A_\infty$
relations is beyond the scope of this work.

Due to the properties of the amplitudes on the torus we can trace back
the holomorphic anomaly of the cylinder diagrams to our computation as
follows. We have seen in the previous section that the disk amplitudes
have well-defined modular properties, and they can be expressed in terms
of the Eisenstein series, $G_{2k}$. Therefore the associated cylinder
amplitudes should also appear in terms of the functions, $G_{2k}$.  In
practice in computing the cylinder diagrams the bosonic zero modes must
be regulated. If the chosen regulator preserves the modularity
in favor of holomorphicity, the second Eisenstein
function $G_2$ is replaced by the
non-holomorphic modular function $\hat G_2$.
$G_2$ is not a modular function of weight $2$ as it 
suffers from a modular anomaly\SchellekensXH.
The function $\hat{G}_2$ is defined by:
\eqn\hatGtwo{\hat G_2(\tau)= G_2(\tau) -{\pi\over\Im\tau} \ . }
If we pragmatically map the holomorphic anomaly of $\hat G_2$ arising in
the cylinder diagrams to the open-string mirror map, $\cN_X$, which
solely contains the Eisenstein function, $G_2$, we obtain the modified
mirror map, $\tilde\cN_X$,
\eqn\NXunhol{\cN_X \rightarrow \tilde\cN_X=
            q^{-{1\over 6}\, {(\cN_u\,u)^2\over\left(2i\Im\tau\right)^2}}\,
            \bar q^{-{1\over 6}\, {(\cN_u\,u)^2\over\left(2i\Im\tau\right)^2}}\,
            \cN_X \ . }
Inserting now the modified mirror map into the mirror
relation~\mirrorrel\ the A-model superpotential becomes:
\eqn\WAenh{\hat\cW = \cN_X^{-3}\,\cW(\tau,\cN_u^{-1}(\hat u-u_0))
           \rightarrow \tilde\pf\,\hat\cW
                   = \tilde\cN_X^{-3}\,\cW(\tau,\cN_u^{-1}(\hat u-u_0)) \ , }
with
\eqn\pflimit{\tilde\pf = q^{{1\over 6}\, {{(\hat u-u_0)}^2\over\left(2i\Im\tau\right)^2}}
     {\bar q}^{{1\over 6}\, {{(\hat u-u_0)}^2\over\left(2i\Im\tau\right)^2}}  \ . }
Now we want to compare this prefactor with the non-holomorphic
factor~\prefact\ on the A-model side.

In the A-model the bulk theory of the torus, $\hat\cT$, depends on a
choice of a base point $(\hat\rho,\bar{\hat\rho})$, where $\hat\rho$ and
$\bar{\hat\rho}$ should be thought of independent variables. Moreover,
the topological A-model, which localizes on worldsheet instantons,
corresponds to the choice $\bar{\hat\rho}\rightarrow+\infty$
\refs{\BershadskyCX,\WittenED}. On the other hand the B-model of the
bulk theory of the torus, $\cT$, does also depend on a choice of base
point $(\tau,\bar\tau)$. The B-model, which is mapped by mirror symmetry
to the A-model at $\bar{\hat\rho}\rightarrow+\infty$, is naturally
identified with the B-model at $\bar\tau\rightarrow+\infty$ \WittenED.
 
For the topological A- and B-model with boundaries we expect also a
dependence on the choice of base point in the moduli space. Therefore,
in addition to the base point in the bulk sector we must also specify a
base point $(\hat u,\bar{\hat u})$ or $(u,\hat u)$ in the boundary
sector.

The prefactor, $\pf$, of eq.~\prefact\ written in terms of $\hat
q=e^{2\pi i \hat\rho}$ and $\bar{\hat q}=e^{-2\pi i \bar{\hat\rho}}$
reads:
\eqn\pfunholom{\pf={\hat q}^{{1\over 6}\, {\hat u^2\over\left(2 i\Im\tau\right)^2}} \,
                   {\hat q}^{-{1\over 6}\, {\bar{\hat u}^2\over\left(2 i\Im\tau\right)^2}} \, 
                   {\bar{\hat q}}^{-{1\over 3}\, {\norm{\hat u}^2\over\left(2 i\Im\tau\right)^2}} \, 
                   {\bar{\hat q}}^{{1\over 3}\, {\hat u^2\over\left(2 i\Im\tau\right)^2}} \ . } 
This agrees with \pflimit\ in terms of the base points $(\hat
u,\bar{\hat u})\rightarrow (\hat u - u_0,0)$ and
$\bar{\hat\rho}\rightarrow\infty$ and $\bar{\hat\tau}\rightarrow\infty$.

\newsec{Conclusions}
In this paper we have computed the topological partition function in 
Landau-Ginzburg B-models on the disk by treating the worldsheet 
superpotential perturbatively. We have argued that the topological disk
partition function computes effective $D$-brane superpotentials.
In two examples we have illustrated that the effective superpotentials 
obtained by this method are compatible with known results in the literature. 
Furthermore, our approach is not limited to the two considered examples, but
instead is also applicable in physically more interesting theories 
such as Landau-Ginzburg models for Calabi-Yau threefolds.

The novel feature of the torus example was the appearance of a marginal 
boundary operator. In comparing with the topological mirror A-model,  
the open-string mirror map became a function of both
closed and open-string moduli, $\tau$ and $u$. An open question is whether 
this open-string mirror map satisfies a partial differential equation 
in $\tau$ and $u$, which can be derived in the B-model from first principles
without making reference to the mirror A-model.
In a similar vein, it would be interesting to find a derivation 
for the heat equation, which is satisfied by the theta functions that make up 
the effective $D$-brane superpotential in the example of the torus.

We have also worked out the relationship between simple 
boundary conditions and matrix factorizations. Matrix factorizations have
a huge gauge invariance which can complicate the analysis as the number of
variables increases. 
By directly imposing these simple boundary conditions, as pursued in this paper,
the gauge redundancies of matrix factorizations do not appear. 
This might be a useful starting point in deriving 
Ward identities which potentially could lead to the type of partial 
differential equations mentioned earlier.

In the two considered examples we have imposed Dirichlet boundary conditions
on all fields in the Landau-Ginzburg model. It is, however, of interest to extend
the presented computation to situations where Neumann boundary conditions do
also arise for some linear combinations of the Landau-Ginzburg fields. This 
occurs, for instance, for the `short' branes in the Landau-Ginzburg model of the 
cubic torus. We hope to pursue this issue in the future.

\goodbreak
\bigskip
\noindent 
{\bf Acknowledgments} 
\medskip \noindent
We would like to thank Ilka Brunner, Stefan Fredenhagen, 
Matthias Gaberdiel, K.S.~Narain and Henning Samtleben for helpful 
comments and useful correspondences. We are especially thankful
to Wolfgang Lerche and Nick Warner for many enlightening discussions
at various stages of this work. SG would also like to thank the Theory
Group at CERN and the ITP at ETH for their hospitality during the course
of this work.

\goodbreak
\bigskip

\appendix{A}{Bulk and boundary supermultiplets}

In our conventions the component expansion of the two-dimensional
$(2,2)$ chiral superfield, $\Phi$, as defined in eq.~\chirality, reads
\eqn\chiralexpansion{
 \Phi = \phi + \sqrt2 \theta^\alpha \psi_\alpha + \theta^\alpha\theta_\alpha F\ . }
Here $\phi$ is the complex bosonic field, $\psi_+$, $\psi_-$, are the
fermionic components and $F$ is the complex auxiliary field of this
chiral multiplet. Let $\epsilon = \frac{\epsilon_+ +
\epsilon_-}{\sqrt2}$ parametrize the unbroken supersymmetry of B-type
boundary conditions. Then one has the following supersymmetry
transformations for the components of the superfields, $\Phi^i$:
\eqn\bulktr{\eqalign{
\delta \phi^i &= -\sqrt2 \epsilon \tau^i \ , \cr
\delta \tau^i &= i \sqrt2 \bar{\epsilon}\partial_x \phi^i \ , \cr
\delta \xi^i &= i \sqrt2  \bar{\epsilon}\partial_y \phi^i + \sqrt2 \epsilon F^i \ , \cr
\delta F^i &= i\sqrt2 \bar{\epsilon}
\left(\partial_y \tau^i - \partial_x \xi^i\right) \ .
}}
In the above equations, we have introduced the combinations
$\tau^i\equiv (\psi_+^i - \psi_-^i)/\sqrt2$ and $\xi^i\equiv (\psi_-^i +
\psi_+^i)/\sqrt2$ as they are more appropriate for the Landau-Ginzburg
model with boundaries. These transformations imply that the $(2,2)$
chiral multiplets, $\Phi^i$, decompose into two multiplets,
$(\Phi_\partial^i,\ \Xi^i)$, under the supersymmetry preserved by the
boundary.  $\Phi^i_\partial$ is a boundary chiral field while $\Xi^i$ is
a multiplet subject to the constraint $D \Xi^i=\sqrt2 F^i$.  In the
absence of a superpotential, we can set $F^i=0$ and then, $\bar{\Xi}$ is
a fermionic (boundary) chiral multiplet.

In the topological model, the generator of the supersymmetry parametrized
by $\bar{\epsilon}$ is conventionally taken to be the BRST operator, $\cQ$.
The zero-form topological observables are thus $\cQ$-closed. In the bulk,
any holomorphic function of $\phi$ is an observable. The one-form and
two-form versions of the bulk operators are given by
\eqn\fops{\eqalign{
V_f^{(0)} &= f(\phi) \ , \cr
V_{\partial_xf}^{(1)} &= \frac{-i}{\sqrt2} (\partial_j f )\ \tau^j \ , \cr
V_{\partial_yf}^{(1)} &= \frac{-i}{\sqrt2} (\partial_j f) \ \xi^j \ , \cr
V_f^{(2)} &= -\frac12 (\partial_i \partial_j f)\ \tau^i \xi^j \ .
}}

When we impose Dirichlet boundary conditions on all the fields in the
Landau-Ginzburg model, the (fermionic) boundary operators for the $A$-type
minimal model are given by
\eqn\Xop{\psi^{(0)}= X \bar{\xi} \ , \quad\quad
         \psi^{(1)}= \frac{~X}{\sqrt2}  \partial_y \bar{\phi} \ , }
and for the cubic torus ($\epsilon^{ijk}$ is the totally antisymmetric tensor
of $SU(3)$)
\eqn\torusops{\eqalign{
\psi^{(0)}=\cmat X^i \bar{\xi}_i \ , \quad\quad\quad\quad&\quad
\psi^{(1)}=\frac{1}{\sqrt2}\,\cmat X^i\,\partial_y\bar{\phi}_i \ , \cr
\Omega^{(0)}= \cmat U\epsilon^{ijk} \bar{\xi}_i \bar{\xi}_j\bar{\xi}_k \ , \quad&\quad
\Omega^{(1)}=\frac{3}{\sqrt2}\,\cmat U\,\epsilon^{ijk} \bar{\xi}_i \bar{\xi}_j\partial_y\bar{\phi}_k \ .
}}
%

\appendix{B}{Combinatorics for the perturbative computation of the torus}
In the computation of diagrams for the torus the combinatorics becomes
easier if we choose to set one bulk insertion and one $\Omega$ insertion
as a zero-form located on the boundary at $+\infty$. This insertion then
provides the three $\bar{\xi}$ zero-modes as well. Thus, we have fixed
the $PSL(2,R)$ invariance and taken care of the fermion zero-modes. This
choice can be carried out for all $\cW_n$ with $n>0$. We will now
consider all possible contractions for this choice when $n=1,3,4$.

In the figures, we will follow the following conventions. The filled dot
indicates a bulk insertion, the unfilled dot represents an $\Omega$
insertion, the cross is a $\psi$ insertion and an open circle with a dot
represents the three fermion zero-modes. Bosonic propagators are
represented by a brown line while fermionic ones are represented by a
blue line. We will indicate by $\gf_n$, the group-theory factor that the
graph represents and $\cfa_n$, the combinatoric factor associated with
the graph, \ie the number of ways that the given graph can be obtained.

\figinsert\cone{$\gf_1 = 3 \kappa_{111} - \frac32 a (\kappa_{123}+\kappa_{132})$}{1in}{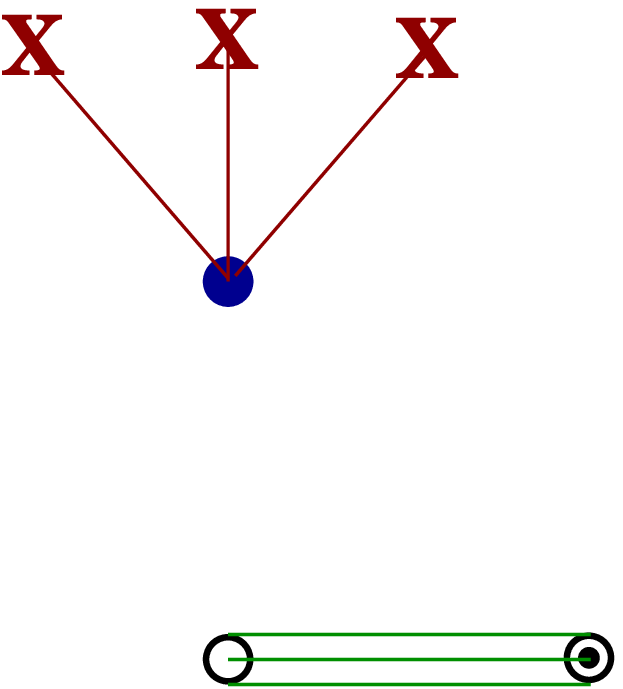}
The combinatoric factor associated with Fig. \cone\ is
$\cfa_1=3!$, which is the number of ways we can carry 
out the $\phi-X$ contractions.

\figinsert\cthree{$\gf_3 = -\frac92 a^2\kappa_{111} +(3 - \frac32 a^3) (\kappa_{123}+
\kappa_{132})$}{1in}{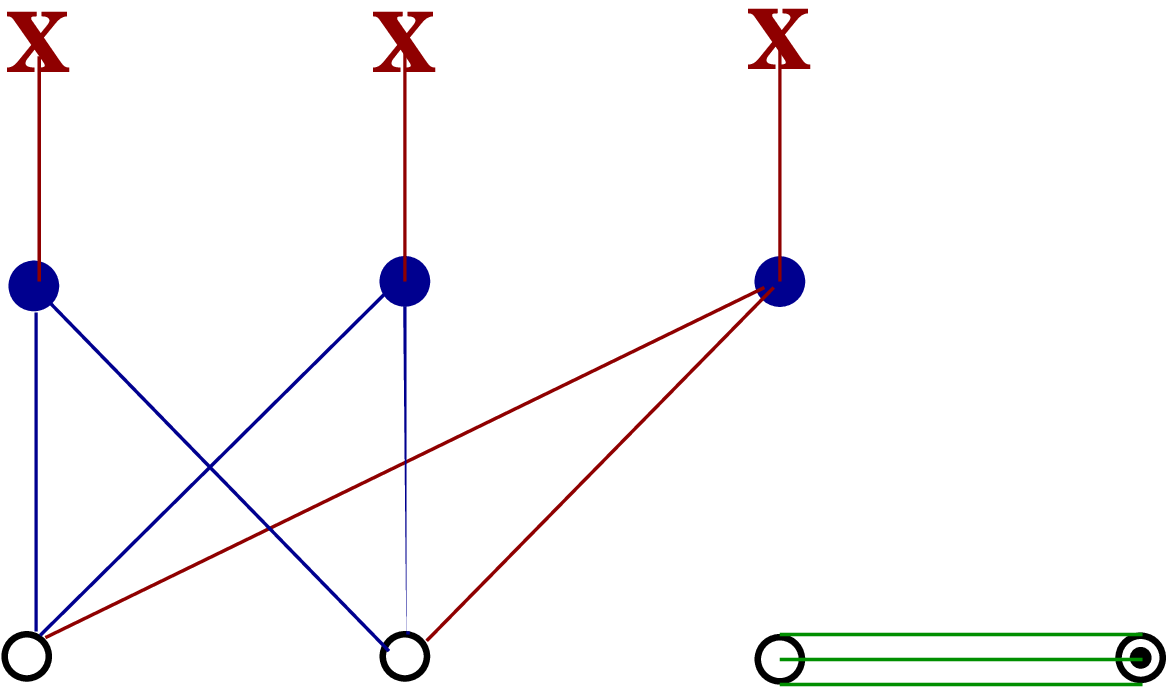}
The combinatoric factor associated with Fig. \cthree\ is
$\cfa_3=(3!)(3!)2^2$. Here the first $3!$ is the number of ways we can
carry out the $\phi-X$ contractions and the second $3!$ is the number of
ways the three $\phi$'s that enter the zero-form bulk insertion
contract. Finally, $2^2$ is the number of possible fermion contractions.
We assume that (we antisymmetrize by hand) $\xi\tau$ fermions that
enters each bulk two-form. So the factor of $2$ arises from the fermion 
contractions with each $\Omega$ insertion. In general, this will contribute 
a factor of $2^{n-1}$ to all graphs. 

We find that when $n=4$, there are two possible graphs which we label
$4a$ and $4b$. In $4a$, the zero-form bulk insertion contracts with one
of the $\psi$-insertions while in the $4b$ it doesn't. Both graphs are
identical except for this difference which shows up as distinct
colorations (coming from the bosonic/fermionic propagators).  So the
combinatoric factors are different.
\figinsert\cfoura{$\gf_{4a} = -(\frac34 a^4 + 6 a) (\kappa_{123}-\kappa_{132})$}{1in}{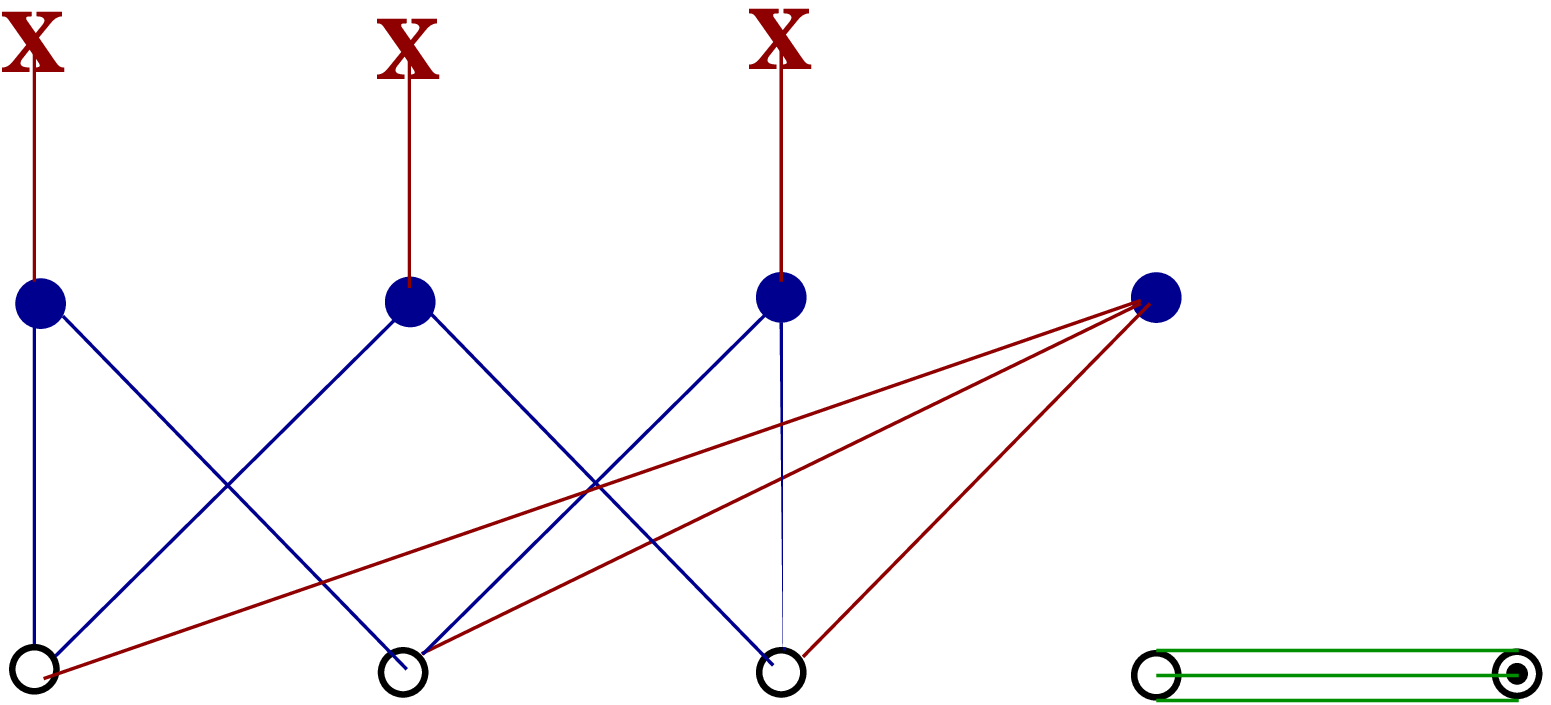}
The associated combinatoric factor is $\cfa_{4a}=(3!)^3 2^3$. Again one
$3!$ arises from the $\phi-X$ contractions, the second from the bulk
zero-form and the last one from the fermion contractions. As explained
earlier, we have $2^{4-1}$ coming from each $\Omega$ one-form.
\figinsert\cfourb{$\gf_{4b} = -(\frac34 a^4 + 6 a) (\kappa_{123}-\kappa_{132})$}{1in}{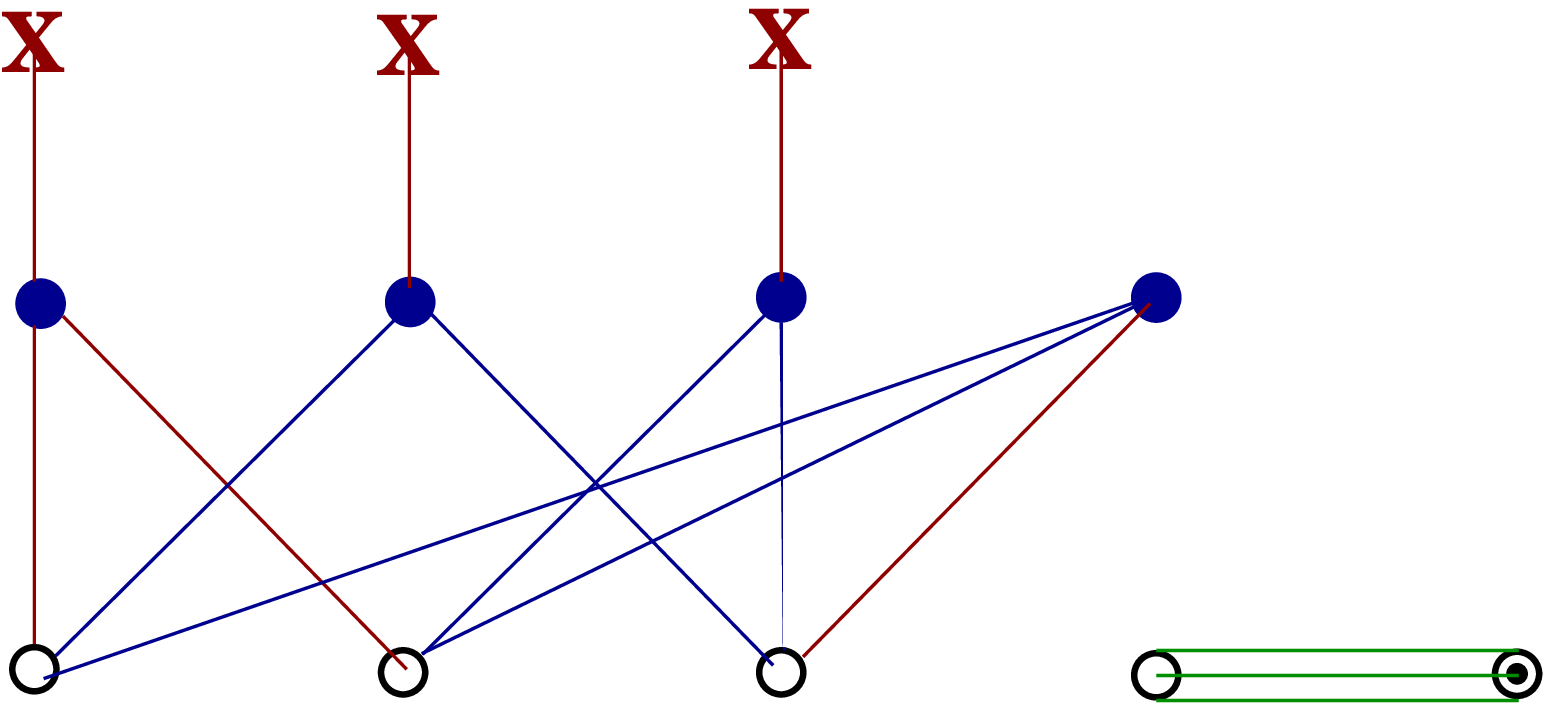}
The associated combinatoric factor is $\cfa_{4b}=(3!)^2 (3) 3! 2^3$.
Here one $3!$ arises from the $\phi-X$ contractions, the second from the
bulk zero-form. The $3$ comes from the number of ways, one of the
$\phi$'s from the bulk-zero form can contract with one of the three
$\Omega$ and the last 3! from the fermion contractions.  
So the contribution is $(\cfa_{4a}\gf_{4a}+\cfa_{4b}\gf_{4b})$.

From $n=5$ onwards the structure is more intricate since the number of
inequivalent graphs increases while their individual complexity is
growing.

\appendix{C}{Integration of the torus correlators} 
The computation of the various free-field correlators requires us to evaluate 
integrals involving Lorentzians $\cL_x$ and $\cL_y$ defined in eq. \Lxy. If all the
integrals were over the real line or real half line, then we could use the residue
theorem to evaluate the integrals because the functions, $\cL_x$ and $\cL_y$, are 
obtained as the real and imaginary parts of a meromorphic function.
However, the required path-ordering implies that this is not
the case for the boundary integrals that we compute.\foot{The path-ordering is 
non-trivial since the Chan-Paton matrices $\cmat{X}^i$ and $\cmat{U}$ do not 
commute for arbitrary $u_i$. They commute for $u_1=u_2=u_3$.} 
So we find that we cannot typically
carry out all integrations. Naively extending the limits of all integrals is
not compatible with the {\it correct} addition of contributions from
different path-orderings that enter the partition function and gives a 
vanishing answer.

There is however, one situation, where we can indeed do the integrals. As we
will see, in a special limit, the integrands simplify to a collection of delta 
functions. This happens when we choose one bulk operator
to be a zero form located at $(x_0,y_0)$ and one boundary operator to
be a zero form located at $x=+\infty$ whereas all other operators are integrated
ones. $PSL(2,\IR)$ invariance implies that the answer must be independent of
$x_0$ and $y_0$. In particular, one can take the limit $y_0\rightarrow 0$.
This is the limit when the unintegrated bulk insertion is taken close to
the boundary of the upper half-plane. Then one finds
\eqn\Lylimit{
\lim_{y_0 \rightarrow 0} \cL_y(w-x_0,y_0) = \pi\ \delta(w-x_0)\ .
}
It is not hard to see that in the limit, $y_0\rightarrow 0$, all 
bulk-boundary contractions involving the bulk zero-form operator reduce 
to such delta functions. Moreover, the resulting product of delta functions 
implies that any boundary operator contracted with the bulk 
zero form gets only a contribution if it is located at $x_0$ on the real line.
Since delta functions are easy to integrate this technique provides a simpler 
way to compute, for instance, the term, $\cW_1$. One may view the bulk zero-form
insertion at $y_0\neq 0$ as a regulator replacing delta functions by Lorentzians
with width $y_0$.

\subsec{The integral for $\cW_3$}
Rather than consider the most general integral, we next consider the integral
that arises in computing the superpotential term, $\cW_3$. Let the bulk zero form 
be located at $(x_0,y_0)$ and the integrated two-form operators at $(x_1,y_1)$
and $(x_2,y_2)$. There are three integrated one form operators, $\psi^{(1)}$,
which we choose to be located at $w_1$, $w_2$ and $w_3$, and two integrated
one-form operators, $\Omega^{(1)}$, whose locations are taken to be at $z_1$
and $z_2$. Then the Lorentzians, which appear in the integrand, are
\eqn\Wthreeintegrand{\eqalign{
&\cL_y(w_1-x_0,y_0) \cL_y(z_1-x_0,y_0) \cL_y(z_2-x_0,y_0) \cr
\times &\cL_y(w_2-x_1,y_1) \Big[ \cL_y(z_1-x_1,y_1) \cL_x(z_2-x_1,y_1) - 
\cL_y(z_2-x_1,y_1) \cL_x(z_1-x_1,y_1) \Big]\cr
\times &\cL_y(w_3-x_2,y_2) \Big[ \cL_y(z_1-x_2,y_2) \cL_x(z_2-x_2,y_2) - 
\cL_y(z_2-x_2,y_2) \cL_x(z_1-x_2,y_2) \Big] \ .
}}
As $y_0$ approaches zero, the three Lorentzians in the first line of 
eq.~\Wthreeintegrand\ tend to be sharply peaked (with width $\sim y_0$) and
become delta function in the limit, $y_0\rightarrow 0$. Note that this implies 
that for small values of $y_0$ the integrand is also sharply peaked at $z_1=z_2$.
Thus we assume that $(z_1-z_2)$ is small and has a fixed sign specified by
the path ordering. Focusing on the second line of eq. \Wthreeintegrand\ and
on carrying out the $x_1$ integration, we find that the answer approaches zero
as $z_1 \rightarrow z_2$ unless $y_1\rightarrow 0$. This behavior can be summarized
as follows
\eqn\Lylimitb{\eqalign{
\int_{-\infty}^\infty dx_1\, & \cL_y(w_2-x_1,y_1) \cr 
\times & \Big[ \cL_y(z_1-x_1,y_1) \cL_x(z_2-x_1,y_1) - 
\cL_y(z_2-x_1,y_1) \cL_x(z_1-x_1,y_1) \Big]\cr
&\ =\ \pi\,\cL_y(w_2-x_1,y_1) \ \cL_x(z_1-z_2,y_1 ) + {\cal O}(z_1-z_2) \ .
}}
Then in the limit $z_1\rightarrow z_2$ one has\foot{We do {\it not} set
${\rm Sign}(0)=0$ (as is often conventional) because our limits are always taken from 
one side.} 
\eqn\Lxlimit{
\cL_x(z_1-z_2,y_1 ) = \frac{\pi}2\ {\rm Sign}(z_1 - z_2)\ \delta(y_1)\ ,
}
where the factor $1\over 2$ reflects the fact that $y_1$ runs over the half line.
Thus, for $y_0$ small but non-zero we obtain that the above
delta function, $\delta(y_1)$, gets replaced by an even function sharply peaked
at $y_1=0$ with width $\sim y_0$. The third line in eq. \Wthreeintegrand\
gives a similar result on carrying out the $x_2$ integration. Therefore
evaluating eq.~\Wthreeintegrand\ in the limit, $y_0\rightarrow 0$, yields 
two identical ${\rm Sign}$ functions, which square to $1$, and a collection of
delta functions, which are easy to integrate.

It is clear that taking the limit, $y_0 \rightarrow 0$, and carrying out the $x_1$
and $x_2$ integrations, the only contribution arises when all the boundary
operators are close to one another.
In other words, the only contribution is a {\it contact term} with each pair
of bulk integrals contributing $\pi^2\over 2$. If there were no path ordering 
to deal with, then each boundary integral would contribute a factor, $\pi$, 
with the end result being $\pi^9\over 4$.

\subsec{Handling the path ordering}
The boundary integrals are path-ordered and we need to carry out the integral
for each ordering separately. However, the fact that they reduce to contact
terms in the limit $y_0\rightarrow 0$ does simplify the analysis. 
First, since the operators, $\psi$, are identical, we can choose to order
the operators such that $z_1 \leq z_2 \leq z_3$. Now all that remains is
to work out the orderings of the boundary preserving operators, $\Omega$. 

Second, note that  all path orderings give the {\it same} answer.
Each such ordering, however, is multiplied by a different $u_i$-dependent factor.
At third order cyclic symmetry implies that there are four different
cyclic invariant combinations of the variables, $u_i$: 
\eqn\cyclicinv{\eqalign{
h_1^{(3)}=u_1^3 + u_2^3 + u_3^3 \ , \quad\quad\quad\quad&\quad
h_2^{(3)}=u_1 u_2^2 + u_2 u_3^2 + u_3 u_1^2 \ , \cr
h_3^{(3)}=u_1^2 u_2 + u_2^2 u_3 + u_3^2 u_1 \ , \quad&\quad
h_4^{(3)}=u_1u_2u_3 \ . }}
It is easy to see that there is one ordering (up to cyclic invariance) that can
give rise to $h_1^{(3)}$, $3$ distinct orderings that give rise to $h_2^{(3)}$ 
(and $h_3^{(3)}$)
and $6$ distinct orderings that give rise to $h_4^{(3)}$. Combining these different
contributions, we see that
\eqn\sumwthree{
\cW_3 \propto (h_1 + 3 h_2 + 3 h_3 + 6 h_4) = (u_1 + u_2 + u_3)^3\ .  }
This shows that only the combination $u\equiv (u_1 + u_2 + u_3)$ can appear (at this
order). Hence, we can choose $u_1=u_2=u_3={u\over 3}$ in order to make $\cmat X$ 
and $\cmat U$ commute. Then, the path-ordering is trivial and the final result 
from the integral is $\pi^9\over 4$.

\subsec{The general case}
The argument of the previous subsections applies also for the two graphs that
appear in evaluating the superpotential term, $\cW_4$. Again only the combination 
$u\equiv (u_1 + u_2 + u_3)$ appears and the integral yields $\pi^{12}\over8$. The
conjecture is that the appearance of the combination, $u$, is true to all orders.
We also suspect that the integrals for the superpotential terms, $\cW_n$, 
are given by $\pi^{3n}\over 2^{n-1}$, \ie a factor, $\pi^2\over2$, from
each bulk integration and a factor, $\pi$, from each boundary integration. 
However, for the term, $\cW_5$, one finds that not all the integrands of the 
contributing graphs simplify to pure contact terms in the limit, $y_0\rightarrow 0$.
Further, the structure of the graphs that appear for higher superpotential terms, 
$\cW_n$, are more complicated and we have not studied them.

\appendix{D}{Computation of the open-string mirror map on the torus}
In this appendix we determine the open-string mirror maps, $\cN_u(\tau)$
and $\cN_X(\tau,u)$. The starting point of this computation is the
mirror-map relation~\mirrorrel, which, as discussed in the main text,
arises from comparing the effective superpotentials of the B- and
A-model, $\cW$ and $\hat\cW$. 

As both correlators, $\Delta_{111}(\tau,u)$ and $\mu_1(\tau,\hat
u)\equiv\hat\Delta_{111}(\tau,\hat u-u_0)$, are odd functions in $u$ and
$\hat u$ respectively, the third power of the mirror map function,
$\cN_X^3$, must be even in the modulus, $u$. Thus it enjoys the
expansion
\eqn\expNu{\cN_X^3(\tau,u)=\sum_{k=0}^{+\infty} c_{2k}(\tau) \ u^{2 k} \ , }
and we need to determine the coefficients, $c_{2k}$, by the
relation~\mirrorrel.

In order to systematically compute the coefficients, $c_{2k}(\tau)$,
which directly specify the mirror map, $\cN_X(\tau,u)$, it is convenient
to first rewrite the functions, $\mu_\ell(\tau,\hat u)$, as power series
in the A-model variable, $\hat u$. The functions, $\mu_\ell(\tau,\hat
u)$, defined by eq.~\defmu, become
\refs{\BrunnerMT,\GovindarajanIM}\foot{The definition of the open-string
modulus, $\hat u$, in ref.~\BrunnerMT\ differs from our open-string
variable, $\hat u$, by a factor of $1/3$. As discussed in
ref.~\GovindarajanIM, the global structure of the open-string moduli
space of the `long' branes~$L_a$ leads in our conventions to the natural
periodicity $\hat u \sim \hat u + 1 \sim \hat u+\tau$.}
\eqn\mudefn{
  \mu_\ell(\tau,\hat u) = e^{\frac{2}{3} i (\ell-1) \pi } 
      \sum_{m\in\IZ} q^{\frac{3}{2} \left(\frac{1-\ell}{3}+m-\frac{1}{2}\right)^2}
       e^{2\pi i \left(\frac{1-\ell}{3}+m-\frac{1}{2}\right) 
                 \left(\hat u-\frac{1}{2}\right)} \ .  }
Note that $\mu_1(\tau,\hat u)$, is odd in $\hat u$, and we also define
the odd and even parts of the functions, $\mu_2(\tau,\hat u)$ and
$\mu_3(\tau,\hat u)$:
\eqn\muoddeven{\eqalign{\mu_2^{\rm o}(\tau,\hat u)
      &={1\over 2}\big(\mu_2(\tau,\hat u)-\mu_2(\tau,-\hat u)\big) 
       ={1\over 2}\big(\mu_3(\tau,\hat u)-\mu_3(\tau,-\hat u)\big) \ , \cr
      \mu_2^{\rm e}(\tau,\hat u)
      &={1\over 2}\big(\mu_2(\tau,\hat u)+\mu_2(\tau,-\hat u)\big) 
       =-{1\over 2}\big(\mu_3(\tau,\hat u)+\mu_3(\tau,-\hat u)\big) \ . }}
After a few steps of algebra we arrive at the power series in the
variable, $\hat u$:
\eqn\muexpand{\eqalign{
  \mu_1(\tau,\hat u) &= \sum_{k=0}^{+\infty} \left({4i\pi\over 3}\right)^k
                  {\pi\over(2k+1)!}\ g^{(k)}(\tau)\ {\hat u}^{2k+1}\ , \cr 
  \mu_2^{\rm o}(\tau,\hat u) &= \sum_{k=0}^{+\infty} \left({4i\pi\over 3}\right)^k
                  {\pi\over(2k+1)!}\ h_{\rm o}^{(k)}(\tau)\ {\hat u}^{2k+1} \ , \cr
  \mu_2^{\rm e}(\tau,\hat u) &= \sum_{k=0}^{+\infty} \left({4i\pi\over 3}\right)^k
                  {i\over(2k)!}\ h_{\rm e}^{(k)}(\tau) \ {\hat u}^{2k} \ . }}
The coefficient functions, $g^{(k)}(\tau)$, $h_{\rm o}^{(k)}(\tau)$ and
$h_{\rm e}^{(k)}(\tau)$, are the $k^{\rm th}$ derivatives of the
functions, $g(\tau)$, $h_{\rm 0}(\tau)$ and $h_{\rm e}(\tau)$, which
turn out to be equal to
\eqn\coeffrel{g(\tau) = 2\, \eta^3(3\tau) \ , \quad\quad
              h_{\rm o}(\tau) = -a(\tau)\, \eta^3(3\tau) \ , \quad\quad
              h_{\rm e}(\tau) = \eta(\tau) \ . }

Now, we have assembled all the ingredients to compute the open-string
mirror maps~\openansatz. At order~$u^0$ we find from the correlator,
$(\kappa_{123}-\kappa_{132})$, in ${\cal W}_0$ with eq.~\CoeffcI
\eqn\czero{c_0(\tau) = {3\, i\, {\cal I}_0 \over \eta(\tau)}={i\over\eta(\tau)} \ , }
whereas at order~$u^1$ we determine with the correlator, $\kappa_{111}$,
in ${\cal W}_1$, the open-string mirror map, $\cN_u(\tau)$, recorded in
eq.~\MirrorNu. There is a consistency check by computing the function,
$\cN_u(\tau)$, as well from the correlator,
$(\kappa_{123}+\kappa_{132})$, in ${\cal W}_1$. This confirms the result
stated in eq.~\MirrorNu.

Since we have already unambiguously determined the open-string mirror
map, $\cN_u(\tau)$, the remaining task is to computed the coefficient
functions, $c_{2k}(\tau)$. This is achieved by comparing in
eq.~\mirrorrel\ the coefficients, $u^{2k}$, of the correlator,
$(\kappa_{123}-\kappa_{132})$, at each order, and we obtain the
recursion relation in terms of the modular
functions~\ModFunc\foot{Recall that $\tilde G_2(\tau)\equiv 0$.} 
\eqn\MirrorRec{\sum_{k=0}^n i\left(3\pi^2\right)^{3(n-k)}
    {\eta^{(n-k)}(\tau)\over (2(n-k))!}\, c_{2k}(\tau) = \tilde G_{2n}(\tau) \ . }
Here we used the expansion~\muexpand\ for the even function, $\mu_2^{\rm
e}$, and the expression~\evenW\ for the B-model superpotential terms,
$\cW_{2k}$. 

With the recursion formula~\MirrorRec\ we are now able to determine the
first few coefficient functions, $c_{2k}(\tau)$:
\eqn\Mirrorc{\eqalign{
    c_0(\tau) &= {i\over\eta(\tau)} \ , \cr
    c_2(\tau) &= {i\over\eta(\tau)}  
                 \left(-{27\pi^6\over 2}{\eta^\prime(\tau)\over\eta(\tau)}\right) \ ,\cr
    c_4(\tau) &= -{i\over\eta(\tau)}\, \tilde G_4(\tau)+
                 {(3\pi^2)^6\,i \over 4} \left({\eta^\prime(\tau)^2\over\eta(\tau)^3}
                 -{\eta^{\prime\prime}(\tau)\over6\,\eta(\tau)^2}\right) \ , \cr
              &\hphantom{=}\ \cdots \ . }}
The structure, which arises from the first few coefficients, $c_{2k}$,
gives rise to the mirror-map function, $\cN_X^3(\tau,u)$,
\eqn\MirrorNXthree{\cN_X^3(\tau,u)={i\over\eta(\tau)}
              \ e^{-{{27\over 8}i\pi^5}\,G_2(\tau)\,u^2+d_4\,G_4(\tau)\,u^4
              +d_6\,G_6(\tau)\,u^6+\cdots} \ . }
Here the coefficients, $d_4, d_6, \ldots$, are determined by the higher
order coefficients, $c_{2k}$. The functions, $G_4$ and $G_6$, are the
Eisenstein functions of modular weight~$4$ and $6$, and the general
expression for the functions, $\cN_X^3(\tau,u)$, is series in terms of
all modular functions.  This form is dictated by comparing yet again the
modular properties of the correlators, ($\kappa_{123}-\kappa_{132}$), of
the B-model to the A-model.  We know that this correlator, as
perturbatively computed in the B-model, is modular invariant. However,
the function, $\mu_2^{\rm e}$, which is the corresponding correlator in
the A-model, is not modular invariant. Hence according to the mirror-map
relation~\mirrorrel\ the product, $\cN_X^3\mu_2^{\rm e}$, must also be
modular invariant.  This, however, is precisely achieved by the stated
mirror-map function~\MirrorNXthree, which contains the compensating
non-modular Eisenstein function, $G_2$.

Except for determining the normalization, $\cN_u$, we have only taken
into account the even terms in $u$ in the effective superpotentials,
$\cW$ and $\hat\cW$.  In other words, we have only used the information
encoded in the correlators, $(\kappa_{123}-\kappa_{132})$. But the
correlators, $(\kappa_{123}+\kappa_{132})$ and $\kappa_{111}$, appearing
only at odd orders in $u$, do also both determine the coefficient
functions, $c_{2k}(\tau)$, recursively. Thus determining the mirror-map,
$\cN_X$, via the correlators, $(\kappa_{123}+\kappa_{132})$ and
$\kappa_{111}$, serves as a highly non-trivial consistency check for the
determined coefficient functions, $c_{2k}$, and for the general method
of computing the potential, $\cW$, perturbatively.

In particular we obtain at third order in $u$ two consistency conditions:
\eqn\relthird{\eqalign{
  0 &= \pi\, {\cal N}_u \, c_2 \, g + {2\pi^2 i\over 9}\, {\cal N}_u^3 \, c_0 \, g'
       -{27\pi^9\over 8\sqrt2} \, a^2\,\qf^{-3} \ , \cr
  0 &=  \pi\, {\cal N}_u \, c_2 \, h_{\rm o} + {2\pi^2 i\over 9}\, {\cal N}_u^3 \, c_0 \, h'_{\rm o}
       +{9\pi^9\over 4\sqrt2} \,\left(1-{1\over 4}\, a^3\right)\,\qf^{-3} \ . }}
The first relation arises from the correlator, $\kappa_{111}$, whereas
the second relation comes from the correlator,
$(\kappa_{123}+\kappa_{132})$. Both identities are fulfilled with the
previously computed quantities, $c_0$, $c_2$ and $\cN_u$. 

Similarly the two relations at order~$u^5$, which are again satisfied 
by the already recursively determined quantities, read:\foot{%
This consistency condition also constraints the coefficient, $\cI_5$, to
$\cI_5 ={\cI_1^5+180 \cI_0^3 \cI_1 \cI_4 \over 15 \cI_0^4 } \ . $ }
\eqn\relfifth{\eqalign{
  0 &= \pi\, {\cal N}_u \, c_4 \, g + {2\pi^2 i\over 9}\, {\cal N}_u^3 \, c_2 \, g'
       -{2\pi^3\over 135}\, {\cal N}_u^5 \, c_0 \, g''
       -3\,{\cal I}_5 \,\left(-{3\over 8} a^4-3a\right)\,\qf^{-5} \ , \cr
  0 &=  \pi\, {\cal N}_u \, c_4 \, h_{\rm o} + {2\pi^2 i\over 9}\, {\cal N}_u^3 \, c_2 \, h'_{\rm o}
       -{2\pi^3\over 135}\, {\cal N}_u^5 \, c_0 \, h_{\rm o}''
       +3\,{\cal I}_5 \,\left({1\over 16}\,a^5-{1\over 2}\, a^2\right)\,\qf^{-5} \ . }}

\vfill\eject
\listrefs
\end